\documentclass[referee,a4paper,12pt,traditabstract]{swsc} 
\usepackage{graphicx}
\usepackage{txfonts}
\usepackage{subfig}
\usepackage{float}
\usepackage{epstopdf}
\usepackage[displaymath,mathlines]{lineno}
\usepackage[authoryear,round]{natbib}
\usepackage[backref]{hyperref}
\usepackage{url}
\usepackage{tikz}
\usetikzlibrary{positioning}

\hypersetup{colorlinks=true,citecolor=cyan,urlcolor=cyan,linkcolor=blue}

\newcommand{\revisiona}[1]{\textcolor{black}{#1}}

\begin{document}


   \title{Leveraging the Mathematics of Shape for Solar Magnetic
     Eruption Prediction}

   \titlerunning{Topology, Geometry and Solar Flare Prediction}

   \authorrunning{Deshmukh et al.}

   \author{V. Deshmukh
          \inst{1}
          \and
          T. E. Berger\inst{2}
          \and
          E. Bradley\inst{3}
          \and
          J. D. Meiss\inst{4}
          }

   \institute{Department of Computer Science, University of Colorado
     430 UCB, Boulder CO
     USA\\ \email{\href{mailto:varad.deshmukh@colorado.edu}{varad.deshmukh@colorado.edu}}
     \and Space Weather Technology, Research, and Education Center
     (SWx-TREC), University of Colorado 429 UCB, Boulder CO USA
     \\ \email{\href{mailto:thomas.berger@colorado.edu}{thomas.berger@colorado.edu}}
     \and Department of Computer Science, University of Colorado 430
     UCB, Boulder CO USA \\and Santa Fe Institute, Santa Fe NM USA
     \\ \email{\href{mailto:lizb@colorado.edu}{lizb@colorado.edu}}
     \and Department of Applied Mathematics, University of Colorado
     526 UCB, Boulder CO USA
     \\ \email{\href{mailto:jdm@colorado.edu}{jdm@colorado.edu}} }

\abstract {Current operational forecasts of solar eruptions are made
  by human experts using a combination of qualitative shape-based
  classification systems and historical data about flaring
  frequencies.  In the past decade, there has been a great deal of
  interest in crafting machine-learning (ML) flare-prediction methods
  to extract underlying patterns from a training set---e.g., a set of
  solar magnetogram images, each characterized by features derived
  from the magnetic field and labeled as to whether it was an eruption
  precursor.  These patterns, captured by various methods (neural
  nets, support vector machines, etc.), can then be used to classify
  new images.  A major challenge with any ML method is the \textit{
    featurization} of the data: pre-processing the raw images to
  extract higher-level properties, such as characteristics of the
  magnetic field, that can streamline the training and use of these
  methods.  It is key to choose features that are informative, from
  the standpoint of the task at hand.  To date, the majority of
  ML-based solar eruption methods have used physics-based magnetic and
  electric field features such as the total unsigned magnetic flux,
  the gradients of the fields, the vertical current density, etc.  In
  this paper, we extend the relevant feature set to include
  characteristics of the magnetic field that are based purely on the
  geometry and topology of 2D magnetogram images and show that this
  improves the prediction accuracy of a neural-net based
  flare-prediction method.}


\maketitle

\section{Introduction}
\label{sec:intro}

Sunspot active regions manifest as large-scale, high-magnitude,
dipolar structures in images of the magnetic field at the surface, or
``photosphere,'' of the sun.  They are the source regions for the
largest solar magnetic eruptions, which produce flares, coronal mass
ejections (CMEs), and energetic particle events that can drive important
space weather events.  Photospheric magnetograms are 2D samples of the
structure of the full 3D solar magnetic field and thus can provide
important clues about the increasing complexity of the magnetic field
in the lead-up to a magnetic eruption---information that can
potentially be leveraged for the purposes of prediction.  While there
has been a great deal of recent work on machine-learning based
algorithms for predicting solar eruptions from magnetogram data, the
{\sl features} used by these algorithms have been predominately
physics-based: taking the curl of the magnetic field to get currents,
computing its gradient, summing up its absolute values, etc.

We propose a new approach to this task: \textit{computing the topology
  and geometry of the structures in 2D magnetograms}.  Instead of
deriving physical quantities from these data sets on a per-pixel
basis, or attempting to model the full 3D coronal magnetic field or
field line connections from the 2D information that is captured in
magnetograms, we formally quantify their structure using the
fundamental mathematics of shape.  We argue that---even though this
procedure ignores the 3D structure or connectivity of the full
field---it enhances the predictive skill of current methods.  Indeed,
it is the current operational practice for human forecasters to
predict solar eruptions by examining sunspot images and/or
magnetograms and using the McIntosh \citep{McIntosh:1990wu} or Hale
\citep{Hale:1919} classification systems to categorize active region
complexity using alphabetical designations. Each category of active
region has a statistical ``24-hour eruption probability'' derived from
historical records that is then reported (following forecaster
adjustments for factors such as rate of flux emergence) as the
eruption forecast for a particular active region \citep{Crown2012}.
Recent statistical methods extend this approach by using historical
flaring rates, together with a Poisson process hypothesis, to develop
more complicated models.
For example, \cite{gallagher2002} use the McIntosh classification to
determine the probabilities of C-, M-, or X-class flares and
\cite{wheatland2004} use a power law distribution of the flare
magnitudes to determine an empirical eruption probability.  However,
these methods have not seen much use in operational forecasting,
primarily because they do not show greater predictive capability than
the historical forecasts that use look-up tables.

Recognizing that the magnetic reconnection that triggers eruptions
takes place in the upper atmosphere (corona) of the sun, attempts have
been made to model the coronal magnetic field using the measured
photospheric field as a boundary condition. The simplest method to
extrapolate the surface field into the corona assumes zero current, so
that the field potential is a solution to the Laplace equation
\citep{Barnes:2006wo, Barnes:2005wv, Wang:1992}.  However, potential
fields cannot store energy---they are lowest energy states---and thus
cannot model the build-up of energy leading to a CME. Other strategies
include Nonlinear Force-Free Field (NLFFF) extrapolations
\citep{Wiegelmann:2012,Aulanier:2005,Demoulin:1996,Priest-Demoulin:1995},
which are known to correlate well with the sites of X-ray flare
emission from eruptions. NLFFF models are presently the most promising
avenue of coronal magnetic field and eruption modeling
\citep{Schrijver:2008tg}, but they have a large number of free
parameters that require extensive manual tuning.  Note that the
photospheric boundary conditions are not sufficient to uniquely
determine the solution \citep{DeRosa:2009ce, Metcalf:2008df,
  Schrijver:2006jw}, and hence the utility of these modeled 3D
structures for forecasting appears to be limited at present.

An alternate approach models the connectivity between opposite
polarity magnetic patches in the photosphere to identify significant
structures such as ``nulls'' and ``separatrix surfaces.'' For example,
\cite{Barnes:2005wv} develop a ``Magnetic Charge Topology'' (MCT) metric
that is used to characterize eruption potential.  \cite{Longcope05}
reviews the application of topology to inferred field line connections
and \citep{Tarr:2012, Tarr:2013} apply these methods to analyze the
eruption potential of active regions.  While these methods are
sophisticated and the structures that they extract are meaningful in
the context of solar eruptions, their computational complexity limits
their operational application.\footnote
{Note that Barnes and Leka have gone on to apply their MCT metric to
a \revisiona{potentially operational} flare prediction algorithm called DAFFS.}

It is important to note that the use of the term ``topology'' in the
methods described in the previous paragraph refers to the study of the
shape and/or connectivity of {\sl 3D field lines} above the
photosphere \revisiona{\cite[e.g.,][]{Longcope05}}.  This contrasts with the more general, mathematical
definition of the term, which refers to the shape and connectivity of
sets of {\sl any} dimension---of which field lines are one instance.
Ideas and techniques from this broader field of topology can be used to
address many other important and potentially meaningful properties of
solar data.  Mathematically, topology
distinguishes sets that cannot be transformed into one another by
continuous maps with continuous inverses (homeomorphisms).  Though
this can obscure much of what is commonly meant by structure, its
roughness can also be a virtue in that it will eliminate distinctions
that could be due to unimportant distortions, e.g., those due to
projection of the sun's spherical surface onto a 2D magnetogram.  {\sl
  Computational} topology, also known as topological data analysis
(TDA), operationalizes this highly abstract framework for use with
real-world data, which may be noisy and poorly sampled.  Though
computing the abstract topology from this type of data is not
feasible, an aspect of shape that can be computed relatively
straightforwardly is \textit{homology}.  In homology, shapes are
distinguished according to their pieces, holes, voids, etc.  This
calculation can be reduced to linear algebra, essentially the
calculation of the dimensions of the ranges and null spaces of certain
matrices \citep{Kaczynski04}.  TDA has been used
in many applications, including coverage of sensor networks
\citep{deSilva07}, structures in natural images \citep{Ghrist08},
neural spike train data \citep{Singh08}, and even the large-scale
structure of the universe \citep{Xu:2018xnz}.  

To compute the topology of a collection of points that samples an
object requires an interpolation scheme to ``fill in'' the gaps
between the points.  The theory of {\sl persistent homology} leverages
interpolation to compute a shape as a function of scale.  The result
is encoded in a plot called a persistence diagram, which can be
further processed to yield useful features. For example, such a
diagram naturally captures \revisiona{how the structures in a
  magnetogram change as flux emerges into an active region during the
  evolution towards a flaring state.}  By contrast, computational
geometry---widely used in computer graphics and computer-aided
design---addresses the problem of extracting purely geometric
information (line segments, polyhedra, etc.) from a data set.
\revisiona{The spatial relationship of the positive and negative
  polarities in an active region, for example---and particularly their
  positioning relative to strong ``neutral lines'' in the photospheric
  field configuration}---are known to be important indicators of
flaring \citep[e.g.,][]{Schrijver}, and computational geometry can
easily capture these properties in formal ways.  We will demonstrate
in \S\ref{sec:shape} that a combination of topological and geometric
analyses can extract meaningful information from a series of
magnetogram examples.

In the past decade, the large increase in magnetogram data afforded by
space missions and advances in data access have shifted the forefront
of flare-prediction research from empirical modeling methods to ``data
analytic'' approaches such as machine
learning. \revisiona{\cite{Camporeale2019} summarizes the state of the
  art in machine learning approaches to space weather applications.}
In ML-based prediction applications, characteristic ``features'' of
the photospheric magnetic field, sometimes combined with features seen
in simultaneous Extreme UltraViolet (EUV) images of the solar corona,
are used in a statistical sense to ``train'' a computational model to
predict the probability of an eruption within a given time period
(usually 24 hours). One example is a support vector machine (SVM)
architecture to perform a binary classification of magnetograms as
flaring or non-flaring
\citep{Bobra:2015fn,Nishizuka:2017,boucheron2015,yang2013,yuan2010}.
\citet{Nishizuka:2017} have also applied decision trees and clustering
to the same task.  Neural networks, which go beyond simple binary
classification by learning complex nonlinear relationships among their
inputs, have also been used to great advantage by
\citet{Nishizuka:2018}.  A variety of other ML algorithms, such as
Bayesian networks \citep{yu2010}, radial basis model networks
\citep{colak2009}, logistic regression \citep{yuan2010},
\revisiona{LASSO regression \citep{Campi2019}, and random forests or
ERTs \citep{Nishizuka:2017, Campi2019}} have also been implemented
for solar flare prediction with some degree of success. \cite{Nishizuka:2017} 
and \cite{Jonas:2018} include Solar Dynamics Observatory (SDO)
Atmospheric Imaging Assembly and EUV image characteristics in addition
to magnetogram data in a fully connected neural network architecture.
\revisiona{\cite{Benvenuto2018} use Fuzzy C-Means---an unsupervised
machine learning method---in combination with some of the mentioned
supervised methods for solar flare prediction.}
\revisiona{Approaches such as \cite{Guerra2018} and
\cite{Kontogiannis2018}, while not machine-learning approaches
themselves, provide statistical tools for evaluating the
engineered features in terms of their potential advantage in machine
learning models.}  \revisiona{Finally, while most of the above
methods focus on developing ML models using engineered features,
recent methods have employed a convolutional neural network (CNN)
approach which automatically extracts features from raw magnetogram
data that are important to predicting flares \citep{Huang2018,
Park2018, Zheng2019}.}

The sophisticated methods described in the previous paragraph
represent the state-of-the-art in ML-based solar eruption prediction.
However, they predominately use physics-based features, and a careful,
quantitative comparison study shows that none of them are
significantly more skilled, and indeed are typically less skilled,
than current human-in-the-loop methods employed at operational
forecasting offices \citep{leka2019a,leka2019b,Barnes:2016bu}.  Our
goal is to improve upon these results using ideas and algorithms from
computational topology and computational geometry to quantify the
complexity in magnetograms and/or sunspot images.  In this study, we
analyze magnetogram images using different flux thresholds---sub- or
super-level sets, in mathematical parlance---and extract structural
signatures that, we conjecture, can be effectively leveraged as
elements in a feature vector for machine-learning methods.  As
evidence in favor of that conjecture, we use high-fidelity vector
magnetic field data from the Helioseismic and Magnetic Imager (HMI)
instrument \citep{Scherrer:2011ji} on the NASA Solar Dynamics
Observatory satellite \citep{Pesnell:2011ik} and show that adding
shape-based signatures to existing feature vectors improves the
24-hour prediction accuracy of a neural-net based method.

This is, to our knowledge, the first time that systematic quantitative
measures of the shape of 2D magnetic structures in the photosphere
have been developed for the purposes of flare prediction.  In a sense,
our approach is a mathematical systemization of the current {\sl ad hoc} 
McIntosh or Hale classification systems.  Though it employs
topology, it is very different from the work described above that
analyzes the magnetic field-line structure.  We focus on the shapes
of {\sl two-dimensional} sets, restricting
our analysis to the photospheric magnetic field
structures. Our goal is to extract formal shape characterizations
that can be leveraged by ML methods to improve flare
prediction.  We are not attempting to model the coronal magnetic field
or determine field line connections between neighboring opposite
polarity structures.
Our approach differs from existing work on geometric
\citep{Schrijver,mcateer2010}
and topological \citep{Knyazeva11,Makarenko14} 
analysis of magnetogram features in that the goal is to derive
features that improve ML-based methods for flare prediction
\revisiona{rather than to discover any one physical property that is
  more or less predictive of flaring}.  We believe 
\revisiona{that this approach has merit since current operational
  flare prediction methods---the McIntosh and/or Hale classification
  systems used by human experts---are fundamentally based on
  active-region shape and geometry.  In addition, \cite{Tarr:2012}
  state that `topological changes' can be shown to precede flaring
  activity in a typical sunspot active region, suggesting that active
  region shape, and its evolution, have a fundamental, meaningful
  connection to the physics of magnetic eruptions.}

The following section goes into more depth on how to formulate and
deploy computational topology and geometry in the context of HMI
magnetograms.  Section \ref{sec:results} presents a study of how
features extracted from magnetograms using those techniques can
improve the prediction accuracy of a specific machine-learning method
for flare prediction.  We conclude in \S\ref{sec:concl}.

\section{Capturing the shapes of active regions}
\label{sec:shape}

\begin{figure}
\begin{center}
\includegraphics[height=0.28\textwidth]{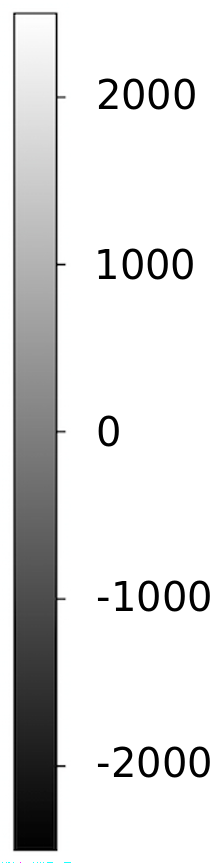}
\includegraphics[height=0.28\textwidth,width=0.28\textwidth]{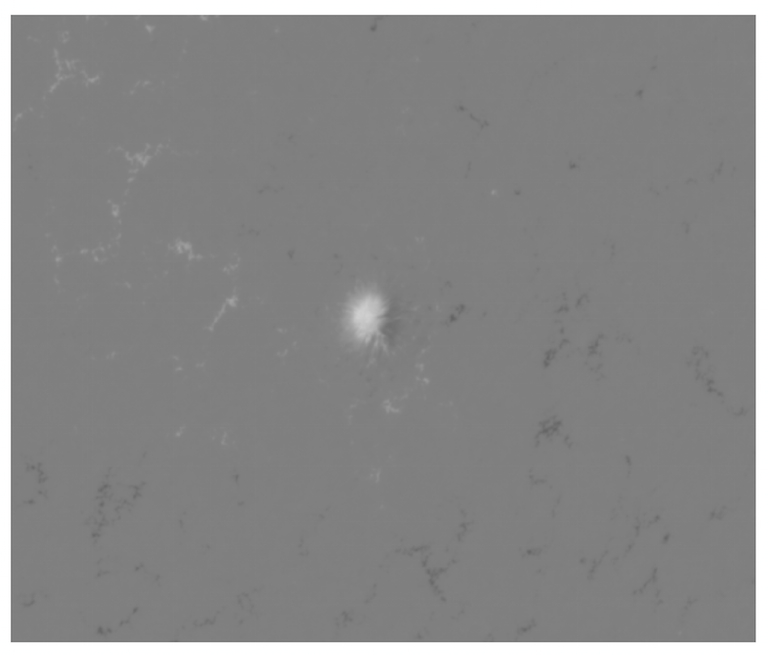}
\hspace*{2mm}
\includegraphics[height=0.28\textwidth,width=0.28\textwidth,angle=180, origin=c]{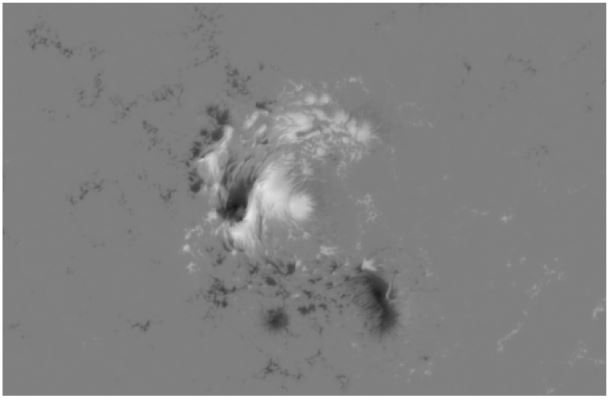}
\hspace*{2mm}
\includegraphics[height=0.28\textwidth,width=0.28\textwidth,angle=180, origin=c]{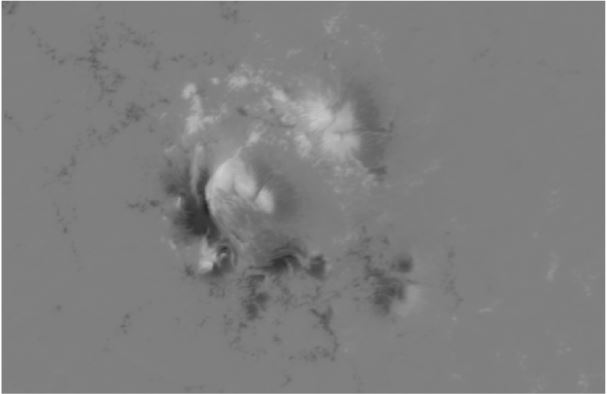}
\centerline{\hspace*{10mm}  (a)\hspace*{45mm} (b) \hspace*{45mm} (c)}
\end{center}
\caption{A series of HMI magnetograms of sunspot \#AR 12673, which
  produced multiple large eruptions as it crossed the disk of the Sun
  in September 2017: (a) at 0000 UT on 9/1, (b) at 0900 UT on 9/5,
  roughly 24 hours before this sunspot produced an X-class solar
  flare, and (c) at 1000 UT on 9/7, around the time of an M-class
  flare.
}
\label{fig:magnetograms}
\end{figure}

Fig.~\ref{fig:magnetograms} shows a series of line-of-sight magnetograms of an
active region before and during an eruptive period.
In panel (a), the active region is newly emerged and is
\revisiona{concentrated into a relatively compact and simple
  configuration.\footnote{\revisiona{We note that since this
      longitudinal flux density measurement is taken close to the
      eastern limb of the Sun, the polarities do not appear to fully
      balance. This is a well-known effect of strong, often
      non-radial, fields in emerging active regions observed near the
      limb.}}
Such initial emergence configurations store little
free energy and are rarely associated with eruptions.}  However, as
more magnetic flux emerges and this active region evolves under the
influence of the plasma flows in the photosphere, it is stretched,
rotated, and sheared into the complex shape shown in panel (b).  While
in this complex configuration, the active region produced a strong
flare that had major impacts on Earth-based radio reception.  The
further development shown in panel (c), later in this sunspot's series
of flaring events, is characterized by intense ``polarity mixing,''
with positive and negative field in close proximity in highly sheared
and stretched shapes. 

Topological Data Analysis or TDA
\citep{{Kaczynski04},{Ghrist08},{Zomorodian12}} is exactly the right
foundation for extracting and codifying the spatial richness of these
images.  Topology is the fundamental mathematics of shape: two sets
are topologically equivalent if they are homeomorphic;\footnote
{That is, they can be mapped to one another by a bi-continuous bijection.}
thus, as is often said, topology does not distinguish between a doughnut and a coffee
cup.  Unfortunately it is not practical to compute the topology of a
set even if one has a complete description of it, let alone when one
only has a finite number of samples.  Topological data analysis was
developed to address these challenges.  It formally quantifies the
shape of a data set according the so-called Betti numbers: the number
of components ($\beta_0$), holes ($\beta_1$), voids ($\beta_2$), etc.,
in the data. 
\revisiona{Note that every topological space has a unique set of Betti numbers,
but these do not completely classify the topology; for example, they would not capture
the twist of a M\"obius strip.}

\begin{figure}
\begin{center}
\includegraphics[width=0.47\textwidth]{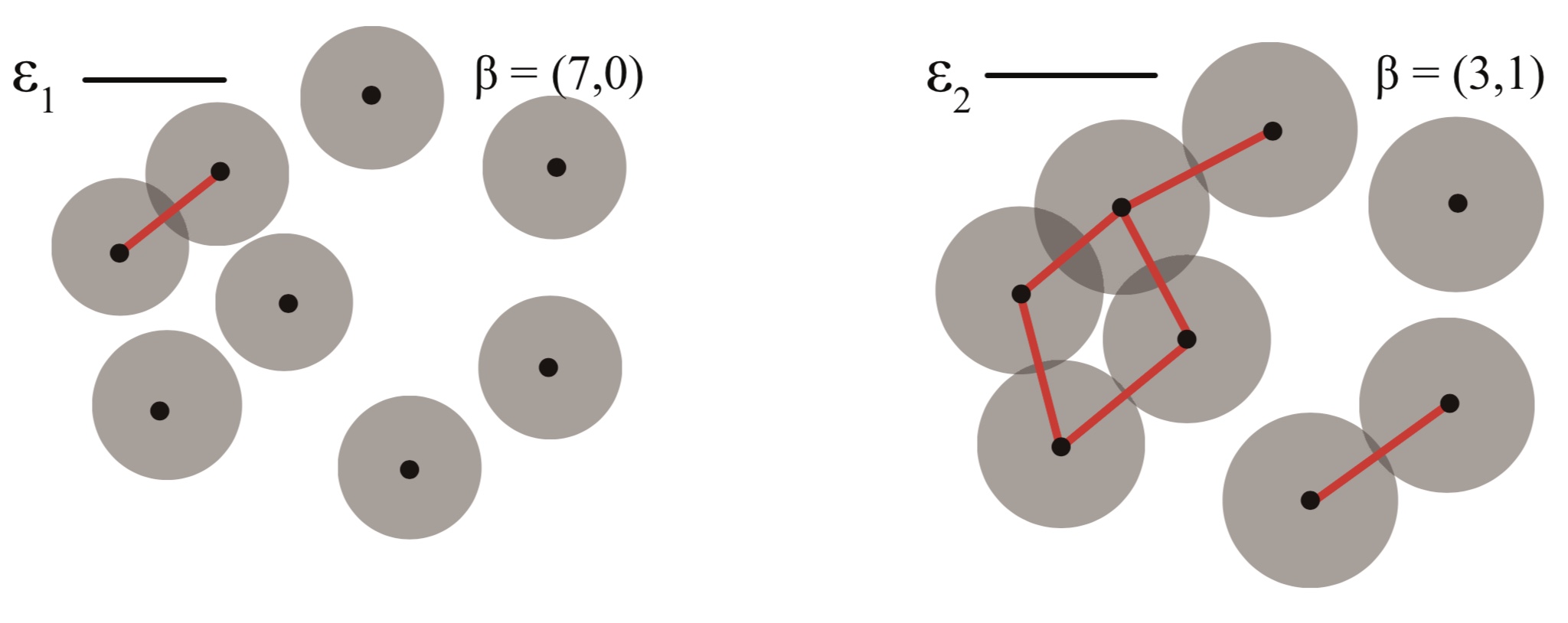} 
\hspace*{5mm}
\includegraphics[width=0.47\textwidth]{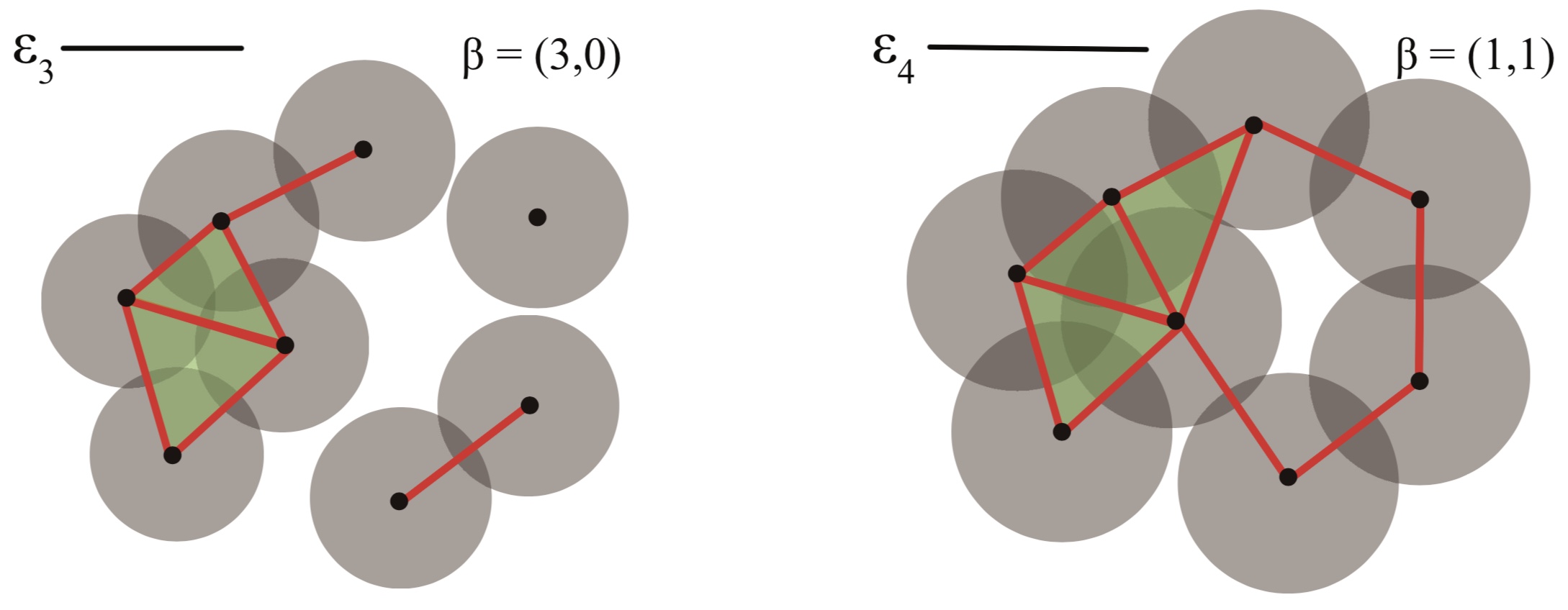} 
\hspace*{2mm}

(a) \hspace*{30mm} (b) \hspace*{30mm} (c) \hspace*{30mm} (d)
\end{center}
\caption{Computational topology: sketch of a data
  set (the black points) enlarged into balls of diameter $\epsilon$,
  for four values of $\epsilon$. The resulting connections, shown as
  lines (red) and triangles (green) form what is called a Rips complex.}
\label{fig:components}
\end{figure}

Of course, a finite collection of points does not really have a
``shape.''  TDA handles this by defining a scale
parameter, for example, creating a manifold from a set of disconnected
points by enlarging each point into a ball, as sketched in
Fig.~\ref{fig:components}.
This leads to the notion of $\epsilon$-connectedness: a pair of points
is treated as connected if they are within a distance $\epsilon$ of
each other.  A set of points connected by a graph with edges of length
no more than $\epsilon$ is called an $\epsilon$ component.  Of course,
when $\epsilon = 0$, each point (black dot in the figure) is a
component.  Conversely, the entire manifold is viewed as connected, from the
standpoint of TDA, if the balls have a sufficiently large radius, as
for the value $\epsilon_4$ in the figure.  \revisiona{The procedure of
  varying scale is familiar from the calculation of fractal dimension:
  one views dimension as reflecting the scale-dependent growth of the
  number of data points in a ball.  However, TDA does not attempt to
  compute shape in the limit $\epsilon \to 0$, as one would for
  fractal dimension; rather it views the change in shape at finite
  values $\epsilon$ as reflecting macroscopic properties of the data
  \citep{Robins00c,Robins98}}.

More generally, the connections give rise to a simplicial complex,
called a Rips\footnote
{An alternative, related complex is the \v{C}ech complex.}
complex; this is essentially a list of these
connections and groups of connections.  Connections between pairs of
points give edges (red lines in Fig.~\ref{fig:components}) that
correspond to 1-simplices. When there is a cycle with three
vertices---i.e., three $\epsilon$-balls pairwise overlap---the
associated triangle (green) is filled in; this is a 2-simplex, etc.
The shape of the complex, and the Betti numbers that describe it, vary
with the scale parameter: when $\epsilon = 0$, $\beta_0$ is equal to
the number of points in the data set and the Rips complex also
contains no higher-dimensional simplices so $\beta_1=0$.  As
$\epsilon$ grows, nearby points are successively joined, causing
components to grow and holes to form. For example for $\epsilon =
\epsilon_2$ in the figure there are three components and one hole, so
$\vec{\beta}=(\beta_0,\beta_1)=(3,1)$, but as $\epsilon$ reaches
$\epsilon_3$, two triangles have formed to fill in the hole and
$\beta_1$ becomes $0$.  Every hole eventually vanishes as the complex
gets filled in, and so for large enough $\epsilon$ there is a single
component with no holes, i.e., $\vec{\beta}=(1,0)$.

\begin{figure}
\begin{center}
\includegraphics[height=0.25\textwidth]{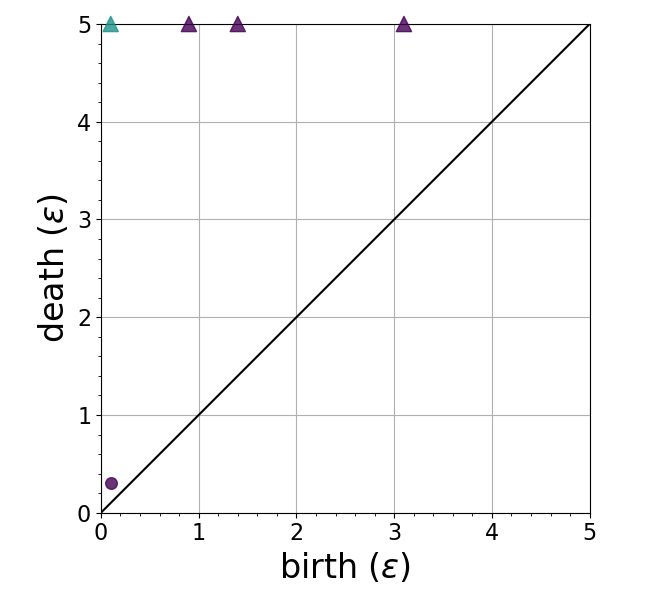}
\includegraphics[height=0.25\textwidth]{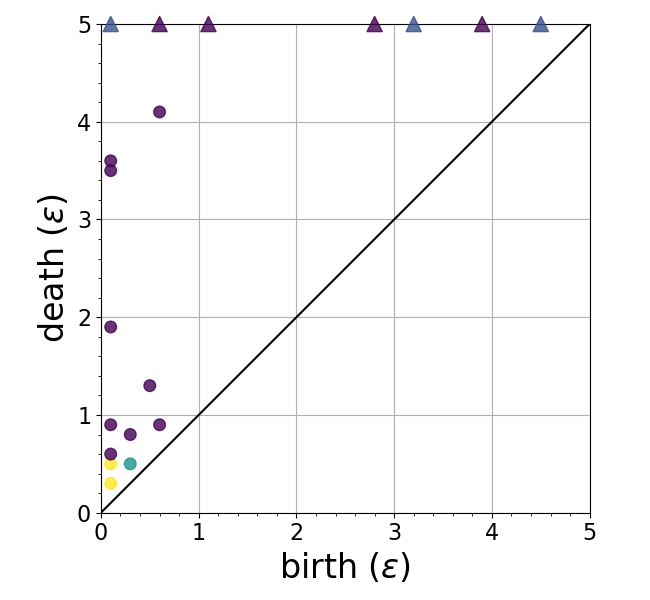}
\includegraphics[height=0.25\textwidth]{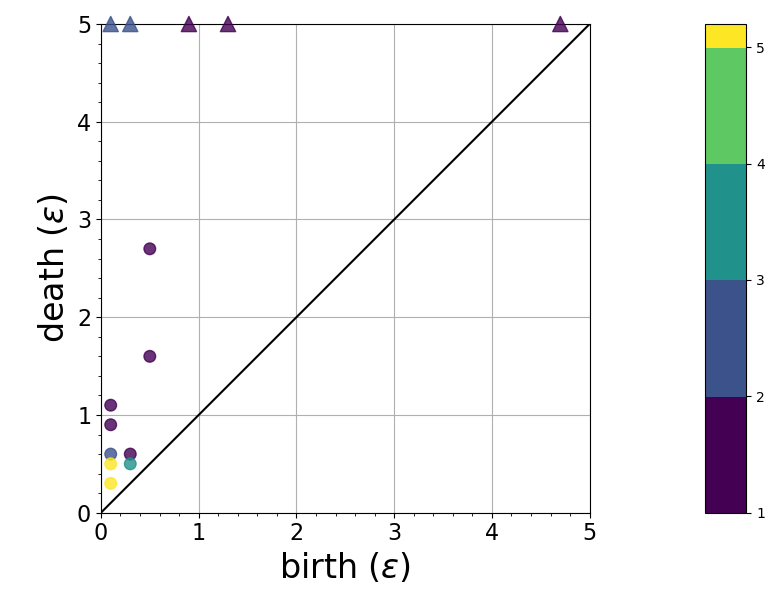}
\centerline{(0000 UT on 9/1/2017)\hspace*{13mm} (0900 UT on
  9/5/2017) \hspace*{13mm} (1000 UT on 9/7/2017) \hspace*{5mm} }
\end{center}
\caption{$\beta_0$ persistence diagrams in a temporal sequence of
  magnetograms of active region \#12673, constructed from the set of
  points (pixels) with positive magnetic field intensities greater
  than 200 Gauss. The co-ordinates of each point $(x,y)$ on a
  persistence diagram represent the birth and death times of a
  $\beta_0$ feature. Note that all the features lie above the $x=y$
  line in the diagrams since a feature cannot die before it is
  born. In terms of number of connected components at various scales,
  these diagrams reveal a clear change in the topology of the field
  structure around the X2.2-class flare that took place at 0910 UT on
  6 September 2017.
}
\label{fig:pds}
\end{figure}

All of this information about the spatial scales of the topological
features in a data set can be captured in a plot called a
\textit{persistence diagram}~\citep{Edelsbrunner00}.  Most
$\epsilon$-components, for example, have birth and death parameter
values---where they appear and disappear, respectively, from the
construction.  A $\beta_0$-persistence diagram has a point at
($\epsilon_{birth}, \epsilon_{death})$ for each component.
Such a diagram is shown in Fig.~\ref{fig:pds} for the series of
magnetograms in Fig.~\ref{fig:magnetograms}.
Components that still exist at the upper end of the calculation
interval---here $\epsilon = 5$ pixels---are represented by triangles
in the figure.
When multiple components have the same $\epsilon_{birth}$ and
$\epsilon_{death}$ values, the color of the icon in Fig.~\ref{fig:pds}
indicates the number of components with that lifespan.  One can also
plot persistence diagrams for the other Betti numbers; the large hole
to the right of center in Fig.~\ref{fig:components}(d), for instance,
would give a point near ($\epsilon_4,\epsilon_5$) on a
$\beta_1$-persistence diagram, where $\epsilon_5$ is the ball radius
that causes all of the $\epsilon$-balls around it to overlap, while
the short-lived hole in panel (b) would be close to the diagonal near
$\epsilon_2$.  This rich representation of information about
the underlying shape that is sampled by a set of data points can, we
conjecture, be effectively leveraged by ML-based flare-prediction
methods. In the case of the magnetograms in
Fig.~\ref{fig:magnetograms}, we have preliminary evidence for this
conjecture: the $\beta_0$-persistence diagrams in Fig.~\ref{fig:pds}
reveal a change in the topology of active region AR \#12673 more than
24 hours before the X-class flare that took place at 0910 UT on 6
September 2017.  \revisiona{To the eye, the change in the overall
number of points on these diagrams is quite obvious; more important
is the number and location of the points that lie far from the
diagonal: i.e., those that persist for wide ranges of the scale
parameter $\epsilon$.  The rich, multi-scale nature of the
information captured in a persistence diagram, and the subjective
nature of some of the associated definitions---viz., the notion of
``far'' from the diagonal---makes it a challenge to develop
effective formal metrics for describing that structure.  This
challenge is a current focus in the TDA research community; see
page~\pageref{page:pd-metrics}.  A full treatment of the associated
solutions is beyond the scope of this paper.}

While the pre-flare change in topology is encouraging, there is
another issue here: the threshold used in these calculations to choose
which pixels in the magnetogram to treat as points in the complex.
Thresholding data into categories involves an awkward choice: rarely
is there a crisp boundary between ``low'' and ``high.''  (Indeed, this
is a large part of the drive to use machine learning in data science.)
While the radial magnetic field strength in an HMI magnetogram ranges
up to about $5000$ Gauss, there is no clear notion of what constitutes
a good threshold value for defining coronal footpoint boundaries.
Moreover, any threshold should be relatively insensitive to the
instrumental noise threshold (on the order of $10\ G$) and the small-scale,
background field (on the order of $100\ G$).

\begin{figure}
\begin{center}
\includegraphics[width=0.25\textwidth]{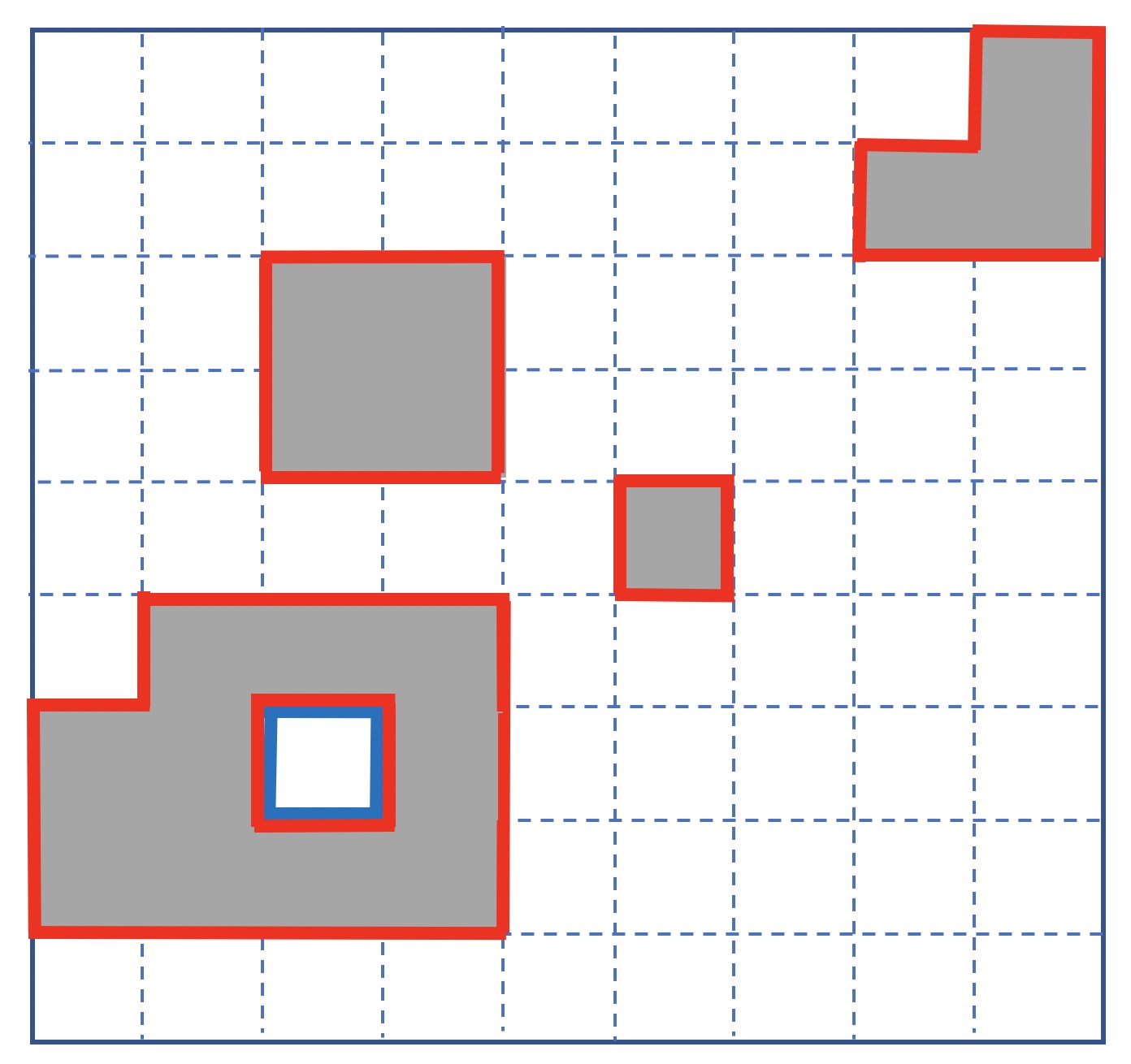}
\end{center}
\caption{In a cubical complex, above-threshold pixels are treated as
  connected iff they are neighbors.  This example has four connected
  components, outlined in red, and one hole, outlined in blue---i.e.,
  $\beta_0=4$ and $\beta_1=1$.}
\label{fig:cubical-schematic}
\end{figure}

\begin{figure}
\begin{center}
\hspace*{-10mm}
\subfloat[Threshold = 1000G]{
\includegraphics[width=0.36\textwidth]{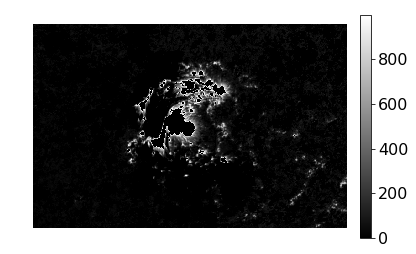}
}
\subfloat[Threshold = 1500G]{
\includegraphics[width=0.36\textwidth]{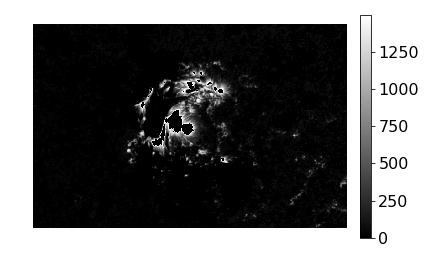}
}
\subfloat[Threshold = 2000G]{
\includegraphics[width=0.36\textwidth]{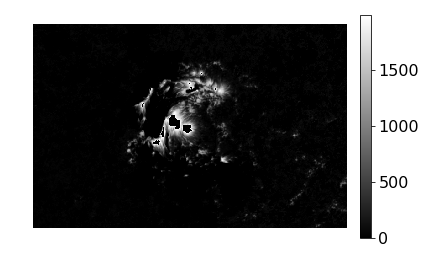}
}
\\
\subfloat[Persistence Diagram]{
\includegraphics[width=0.35\textwidth]{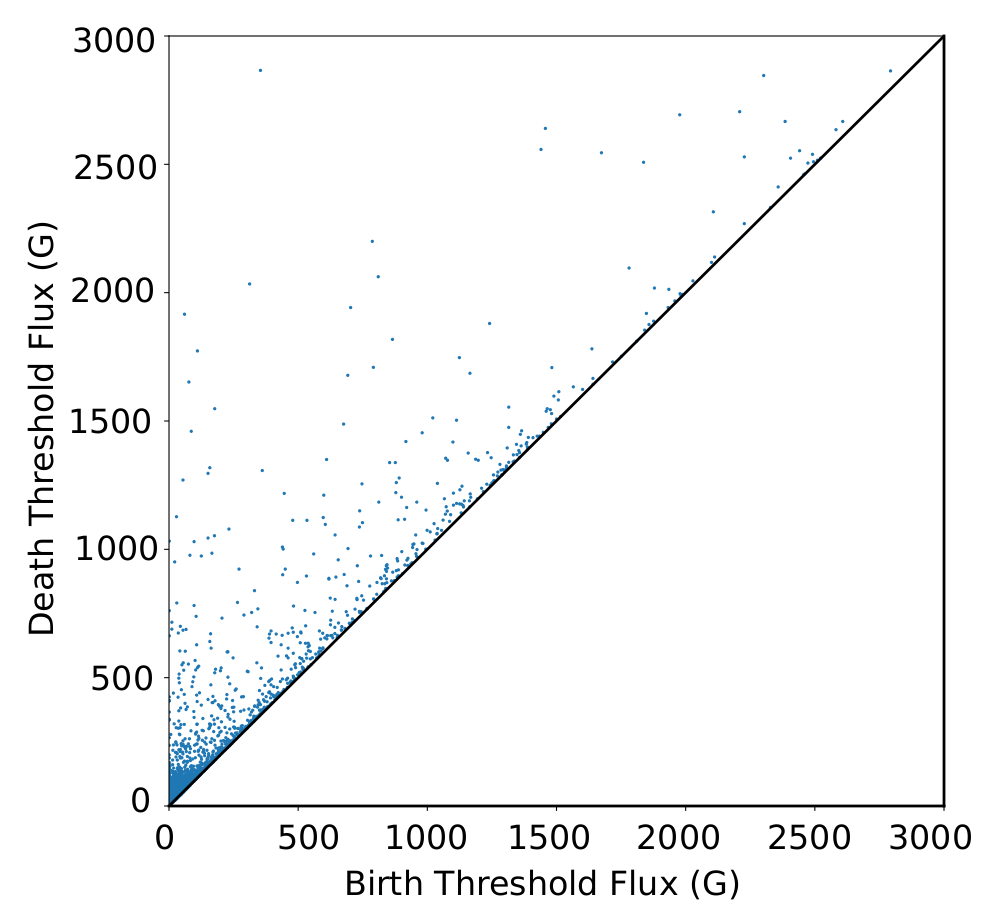}
}
\end{center}
\caption{Cubical complexes for AR \#12673 at three example threshold
  values and the full associated $\beta_1$ persistence
  diagram for a range of positive magnetic field thresholds.}
\label{fig:cubical}
\end{figure}

The proof-of-concept persistence diagrams in Fig.~\ref{fig:pds} were
constructed from thresholded images containing only pixels with
magnetic field intensity greater than $200\ G$; i.e., a super-level set
of the intensity.  Instead of some arbitrary choice, it would be
preferable to view the threshold itself as a parameter to vary,
thereby obtaining a sequence of super-level sets.
An elegant alternative is to simply use the threshold value, rather
than the spatial scale, as the persistence parameter.  To eliminate
the spatial scale parameter $\epsilon$ of the Rips complex that is
appropriate for arbitrary point clouds like that shown in in
Fig.~\ref{fig:components}, we note that magnetograms are pixed-based
data.  The simplest construction of a complex in this case uses a
{\sl cubical complex}.  Here, one simply treats two pixels as
connected if they share an edge or a vertex and each
represents a flux level that is below the chosen threshold (i.e., sub-level thresholding).
Fig.~\ref{fig:cubical-schematic} shows a schematic of such a
construction;
Fig.~\ref{fig:cubical}(a-c) shows cubical complexes constructed for
three threshold values for AR \#12673.  Note how the holes in
Fig.~\ref{fig:cubical}(a-c) form and then fill in as the threshold
changes.
This gives a different view on persistent homology and a second kind
of persistence diagram: one with the threshold value, rather than the
connectedness scale, on the $x$ and $y$ axes, as shown in
Fig.~\ref{fig:cubical}(d).  This persistence diagram is far more detailed
than the ones in Fig.~\ref{fig:pds} because it captures structures at
a {\sl range} of thresholds, and thus is more affected by the
complexity of the structure of this active region.  The patterns in
this plot---the large number of short-lived holes near the diagonal
and the long-lived holes further above and to the left---are an
accurate formal representation of that complexity\footnote
{Some persistence diagram analysis techniques discount or discard points
  near the diagonal, as they are more sensitive to noise and
  pixelation of the data.}.

Persistence diagrams are powerful tools, but topology alone is not a
complete tool for characterizing the shape of structures in a
magnetogram.  Simply identifying the number of high-flux regions, for
instance, says nothing about their sizes or proximities.  A pair of
converging high-flux regions looks the same, from the standpoint of
topology until they actually touch\footnote{\revisiona{From the standpoint of
  {\sl computational} topology, the definition of ``touch'' depends on
  the scale parameter $\epsilon$.}}---and topology cannot measure
quantitative features like the total magnetic flux in such a region.
In order to capture these important, and potentially predictive,
properties so that they can be used by ML algorithms, we extract
geometric information as well by computing the sizes of the
high-flux regions and the distances between them, summing up the flux
inside them, finding their centroids, and the like.  Computational
geometry algorithms \citep{Forrest187,preparata} are widely used for
these kinds of calculations across many fields of science and
engineering, including astronomy---e.g., shape reconstruction for
asteroids \citep{Devogele} and galaxy distribution analysis
\citep{Bhavsar1996}.

Our approach to distilling informative, discriminating features for
each active region out of solar magnetogram data builds on all of
these foundations.  Any or all of the topological and geometric
properties described in this section might be meaningful predictors of
flaring.  Moreover, it is not only the topology and geometry of the positive
and negative regions of the field that are indicative, but also their
topological and geometric relationships to one another; so we also
explore composite features, as described in the following section.

\section{Results}
\label{sec:results}

As demonstration of the utility of shape-based features in
machine-learning methods for solar flare prediction, we chose to work
with an Artificial Neural Network or ANN \citep{Haykin1998}, a
machine-learning approach to fitting a repertoire of nonlinear
functions to the data.  ANNs, also known as multilayer perceptron
architectures, are both flexible and powerful; they can \revisiona{generally}
model more-complex nonlinear functions than regression networks like
SVMs or decision trees.  An ANN is built by stacking together layers
of nonlinear elements, known as {\sl neurons}, with weighted
interconnections between consecutive layers.  The input layer takes
the magnetogram features and feeds them into the stack. A
layer-by-layer, feed-forward propagation of activations finally results in
a prediction at the output layer---in this case, a binary
value that classifies the magnetogram as a precursor to a flare (or
not), i.e., whether that magnetogram is followed by an eruption above
the X-ray class of M-flares within the next 24 hours.  This
classification of magnetograms is equivalent to the “M1.0+/0/24” event
designation in \cite{leka2019a}.

The features used in this study were derived from a set of HMI SHARP
region magnetograms, described in \S\ref{sec:data}.  We preprocessed
each using the techniques described in the previous section to
extract a suite of geometry- and topology-based features.  These
feature sets, described in \S\ref{sec:features}, were tested both as
the sole input vectors to \revisiona{a many-layered fully-connected deep neural network system 
and in combination with the physics-based feature vectors used in current ML
models. }

The design of a ANN requires setting various parameters, including the
number of layers and the number of neurons in the ``hidden layers''
between the input and the output layers; these details are discussed
in Appendix~\ref{sec:appendix_ann}.  The process of {\sl training} the
ANN involves tuning the weights in the layer interconnections so that
the ANN generates the correct labels for the data, given values for
all of the input features. To implement this, we divided the
data into a training set and a testing set; see
\S\ref{sec:training}.  The training data were used by standard
optimization algorithms to tune the weights.  We repeated this
using various combinations of the feature sets and evaluated the
resulting prediction accuracy using the True Skill Score on the
testing set; results are reported in \S\ref{sec:evaluation}.

\subsection{Data}
\label{sec:data}

The Helioseismic and Magnetic Imager (HMI) instrument
on the NASA Solar Dynamics Observatory satellite views the entire solar disk at a nominal data cadence of 12 minutes, and has captured every active region on the
Earth-facing side of the Sun since its launch in 2010.  The HMI
dataset contains {\em Space Weather HMI Active Region Patches}
(SHARPs) that provide cut-out regions around each of these active
regions, each with vector magnetic field and other derived quantities
calculated for all pixels in the region
\citep{Bobra:2014dn}
as it rotates across the Earth-facing hemisphere of the Sun.
For this study, we used only $B_r$, the radial field component, of SHARPs
images along with the associated metadata taken by the HMI instrument between January 2010 and December 2016.
\revisiona{The $B_r$ component image data is available in the JSOC {\tt hmi.sharp\_cea\_720s} dataset, where the magnetic field vector $B$ is remapped to a Lambert Cylindrical Equal-Area (CEA) projection, and decomposed into the components $B_r$, $B_\theta$ and $B_\phi$.}
This segment of the SHARPs dataset contains about 2.6 million data
records, each approximately 2 MB in size, totaling 5 TB of data.
These active regions are known to have produced around $1250$
M1.0+ flares within 24 hours of the image time---see, for example, Tbl.~1 in \citep{Schrijver:2016cs}.
We downsampled the $B_r$ dataset to a one-hour cadence (i.e., taking every
5th magnetogram),
then used the the NOAA Geostationary Operational Environment Satellite (GOES) X-ray Spectrometer (XRS) flare catalog\footnote
{\scriptsize \tt www.ngdc.noaa.gov/stp/space-weather/solar-data/solar-features/solar-flares/x-rays/goes/xrs/}
to label each one as to whether or not the associated active region
produced an M1.0+ flare in the 24 hours following the time
of the sample.
\revisiona{Next, we discarded all the magnetogram images that contained
  invalid pixel data (NaN values).}  The resulting data set included
\revisiona{$3691$} active regions, of which \revisiona{$141$} produced
at least one M1.0+ flare as they crossed the Sun's disk and
\revisiona{$3550$} did not.  This corresponded to
\revisiona{$438,539$} total magnetograms, of which \revisiona{$5538$}
and \revisiona{$432821$}, respectively, were labeled as flaring and
non-flaring.  A large positive/negative imbalance like this is a
major challenge for any machine-learning algorithm, as described further below.

\subsection{Features}
\label{sec:features}

Values for a number of physics-based features, like the total unsigned
flux in the active regions and the vertical current helicity, are
included in the image metadata for all SHARPs data products; see
Tbl.~\ref{tab:sharps-features} for a full list of these quantities
and {\tt \small jsoc.stanford.edu/doc/data/hmi/sharp/sharp.htm} or
\cite{Bobra:2014dn} for details about these values and the associated
calculations.
\begin{table}
\begin{center}
\begin{tabular}{| l | l | l |}
\hline
Acronym & Description & Units\\
\hline
LAT\_FWT & Latitude of the flux-weighted center of active pixels & degrees \\
LAT\_FWT & Latitude of the flux-weighted center of active pixels & degrees \\
AREA\_ACR & Line-of-sight field active pixel area & micro hemispheres \\
USFLUX &  Total unsigned flux & $Mx$ \\
MEANGAM &  Mean inclination angle, gamma & $degrees$ \\
MEANGBT &  Mean value of the total field gradient & $G/Mm$ \\
MEANGBZ &  Mean value of the vertical field gradient & $G/Mm$ \\
MEANGBH &  Mean value of the horizontal field gradient & $G/Mm$ \\
MEANJZD &  Mean vertical current density & $mA/m^2$ \\
TOTUSJZ &  Total unsigned vertical current & $A$ \\
MEANALP &  Total twist parameter, alpha & $1/Mm$ \\
MEANJZH &  Mean current helicity & $G^2/m$ \\
TOTUSJH &  Total unsigned current helicity & $G^2/m$ \\
ABSNJZH &  Absolute value of the net current helicity&  $G^2/m$ \\
SAVNCPP &  Sum of the absolute value of the net currents per polarity & $A$ \\
MEANPOT &  Mean photospheric excess magnetic energy density & $ergs/cm^3$ \\
TOTPOT &  Total photospheric magnetic energy density & $ergs/cm^3$ \\
MEANSHR &  Mean shear angle (measured using $B_{total}$) & $degrees$ \\
SHRGT45 &  Percentage of pixels with a mean shear angle greater than 45 degrees & $percent$ \\
\hline
\end{tabular}
\end{center}
\caption{SHARPs magnetic field features: values for these 19 features,
  as well as estimates of the errors in those calculations, are
  provided for each magnetogram in the SHARPs database. Abbreviations: $Mx$ is Maxwells, $G$ is
  Gauss, $Mm$ is megameters, and $A$ is Amperes.}
\label{tab:sharps-features}
\end{table}
This feature set---the standard in ML-based flare prediction work---is
the base case for our comparison experiments.
  
Our procedure for computing geometry- and topology-based features from
each magnetogram was as follows.  To remove ``topological'' noise, we
first filtered out pixels whose magnetic flux magnitude was below
$200\ G$, then aggregated the resulting pixels into clusters.  We
computed the number and area of these clusters, then discarded all
clusters whose area was than 10\% of the area of the maximum cluster.
We performed these operations separately for the positive ($>200\ G$)
and negative ($<-200\ G$) fields.  We then computed an {\sl interaction
  factor (IF)} between all positive/negative pairs; this quantity is
defined a manner similar to the so-called Ising Energy used by
\citet{Florios2018} (first introduced in \cite{Ahmed2010}):
\[IF=\frac{B_{pos} \times B_{neg}}{r_{min}^2}\]
\noindent where $B_{pos}$ and $B_{neg}$ are the sums of the flux over
the respective components and $r_{min}$ is the smallest distance
between them.\footnote
{\revisiona{\citeauthor{Florios2018} compute Ising Energy by
    aggregating over all pairs of individual monopoles across the two
    polarity regions: $-\sum_{ij} \frac{S_{i}S_{j}}{d_{ij}^{2}}$ ($S_{i} = \pm1$ 
    for positive/negative pairs). On the other hand, IF
    is computed using the summed
    fluxes over the two polarity regions and the smallest
    distance between them.}}
  We then chose the pair with the highest $IF$ value and derived a
  number of secondary features from that ``most interacting pair''
  (MIP). These features are listed in
  Tbl.~\ref{tab:geom-features}. \revisiona{For magnetogram images
    with only one sign of polarity, we assigned all features involving
    the missing polarity a default value of 0, including those that
    contain the distance terms between opposite polarities.}
\begin{table}
\begin{center}
\begin{tabular}{| l | l | c |}
\hline
Name & Units & \# of features \\
\hline
Total number of positive (negative) clusters & $integer$ & 2\\
Size of largest positive (negative) clusters & $arcseconds$ & 2\\
Interaction factor of MIP & $G^{2}/arcseconds^2$ & 1\\
COM distance between positive and negative elements of MIP & $arcseconds$ & 1\\
Smallest distance between elements of MIP & $arcseconds$ & 1 \\
COM distance to smallest distance ratio & $arcseconds$ & 1 \\
Total magnetic flux of each element of MIP & $G$ & 2 \\
Size of each element of MIP & $arcseconds^2$ & 2 \\
Total magnetic flux per unit area of each element of MIP & $G/arcseconds^2$ & 2\\
Total magnetic flux of largest elements in the magnetogram & $G$ & 2\\
\hline
\end{tabular}
\end{center}
\caption{Geometry-based features: 16 total for each magnetogram. All
  distances were measured in terms of pixels for SDO HMI magnetogram
  images with a pixel resolution of one arcsecond (1 arcsecond = 725
  km).}
\label{tab:geom-features}
\end{table}

To compute topological features from these data, we constructed
cubical complexes for each magnetogram across the range of magnetic
flux magnitude thresholds between \revisiona{263.15-5000 $G$}
divided into \revisiona{ten equally spaced magnetic flux values (both 
  263.15 $G$ and 5000 $G$
  inclusive): $\lbrace263.16\ G, 789.47\ G,
  1315.79\ G,\ \ldots\ 5000\ G\rbrace$}.\footnote
  {The specific threshold
  values come from dividing the interval [-5000 $G$, 5000 $G$] into 20
  equally spaced flux values, since we are performing the analysis for
  positive and negative magnetic fluxes as explained further.}
This range covers all relevant flux levels from quiet Sun network to
sunspot umbral cores. We repeated this operation separately for the
positive and negative fields, yielding a total of $10 + 10 = 20$
cubical complexes.  We then used those complexes to construct
$\beta_1$ persistence diagrams, like that in
Fig.~\ref{fig:cubical}(d), that capture the threshold values at which
each hole in that field is born and dies.  Finally, we transformed
these 2D diagrams to vectors that can be more-effectively leveraged by
machine-learning methods, one vector each for the positive and the
negative magnetic flux values.

In the past few years, there has been a growing literature on the problem of
featurizing persistence diagrams.  There are two classes of
approaches: finite-dimensional embeddings or kernel-based methods.
Persistence landscapes \citep{Bubenik2015}, persistence images
\citep{Adams2017} and persistence silhouettes \citep{Chazal2014} are
some examples of the first approach that fit template functions to the
diagrams.  Kernel-based methods \citep{Bubenik2015, Reininghaus2015,
  Kusano2016, Carriere2017, Le2018} use generalized scalar products
that transform the diagrams implicitly into infinite-dimensional
Hilbert Spaces.  While these methods are useful in defining meaningful
relationships between two persistence diagrams, their inherent
compression can lead to a loss of information. To address this,
\cite{Carriere2019} propose a layer for neural network architectures
that encodes most of these vector representations using a set of
generalized point-wise transformation functions.
\label{page:pd-metrics}
Here, we chose a very simple version of the first class of approaches,
counting the number of ``live'' holes at each value of the threshold
flux.
Since we have ten cubical complexes for each polarity of the field,
this produces two vectors with ten entries each.  We concatenate these
together to create a single vector of length 20---the topology-based
features for our study. \revisiona{Tbl.~\ref{tab:feature-sets}
  summarizes the three basic feature sets of interest evaluated in
  this study.}

\begin{table}
	\begin{center}
		\begin{tabular}{| c | c |}
			\hline
			Feature Set & Number of features \\
			\hline
			SHARPs (baseline) & 19 \\
			Geometry-based & 16 \\ 
			Topology-based & 20 \\
			\hline
		\end{tabular}
	\end{center}
\caption{The three basic feature sets evaluated in this study. The performance
	of a combination of some of these feature sets is also studied, as 
	described in Section~\ref{sec:evaluation}.}
\label{tab:feature-sets}
\end{table}

\subsection{Training}
\label{sec:training}

We followed standard procedures to train and evaluate the ANN model,
beginning by splitting the dataset into training and testing sets.
During the training phase, the output labels produced by the ANN,
working from the features derived from each magnetogram in the
training set, were compared with the true labels for that magnetogram.
The difference was then propagated backwards through the network to
update the weights of the interconnections between the layers.  Via
multiple passes through the entire dataset (``epochs''), the weights
were updated until the ANN learned a set of weights that sufficiently
fit the data.  Once the training phase was completed, the weights of
the model were frozen permanently and the ANN could be used to make
predictions on the testing data.  Details and citations for all of
the steps in this process can be found in
Appendix~\ref{sec:appendix_ann}.

The normal approach to splitting the data into training and testing
sets---random shuffling---is not appropriate in this application.
Instead, we split the magnetogram images dataset into training and
testing sets {\sl based on their SHARPs IDs}. \revisiona{That is, we
  randomly chose 70\% of the SHARPs regions and placed all feature
  vectors extracted from the associated magnetograms in the training
  set. The feature vectors extracted from magnetograms from the
  remaining 30\% made up the testing set.}  This ensured that the
\revisiona{features} used in evaluating the ANN did not have any
similar counterparts that were used by the model during the training
phase, thereby avoiding artificial boosting of prediction scores. 
\revisiona{We repeated this random shuffling procedure with 10 different 
random seeds to generate statistical results across 10 different
training/testing sets.}

\subsection{Evaluation and Discussion}
\label{sec:evaluation}

\begin{figure}
	\begin{center}
		\hspace*{-10mm}
		\subfloat[Geometry Experiments]{
			\includegraphics[width=0.36\textwidth]{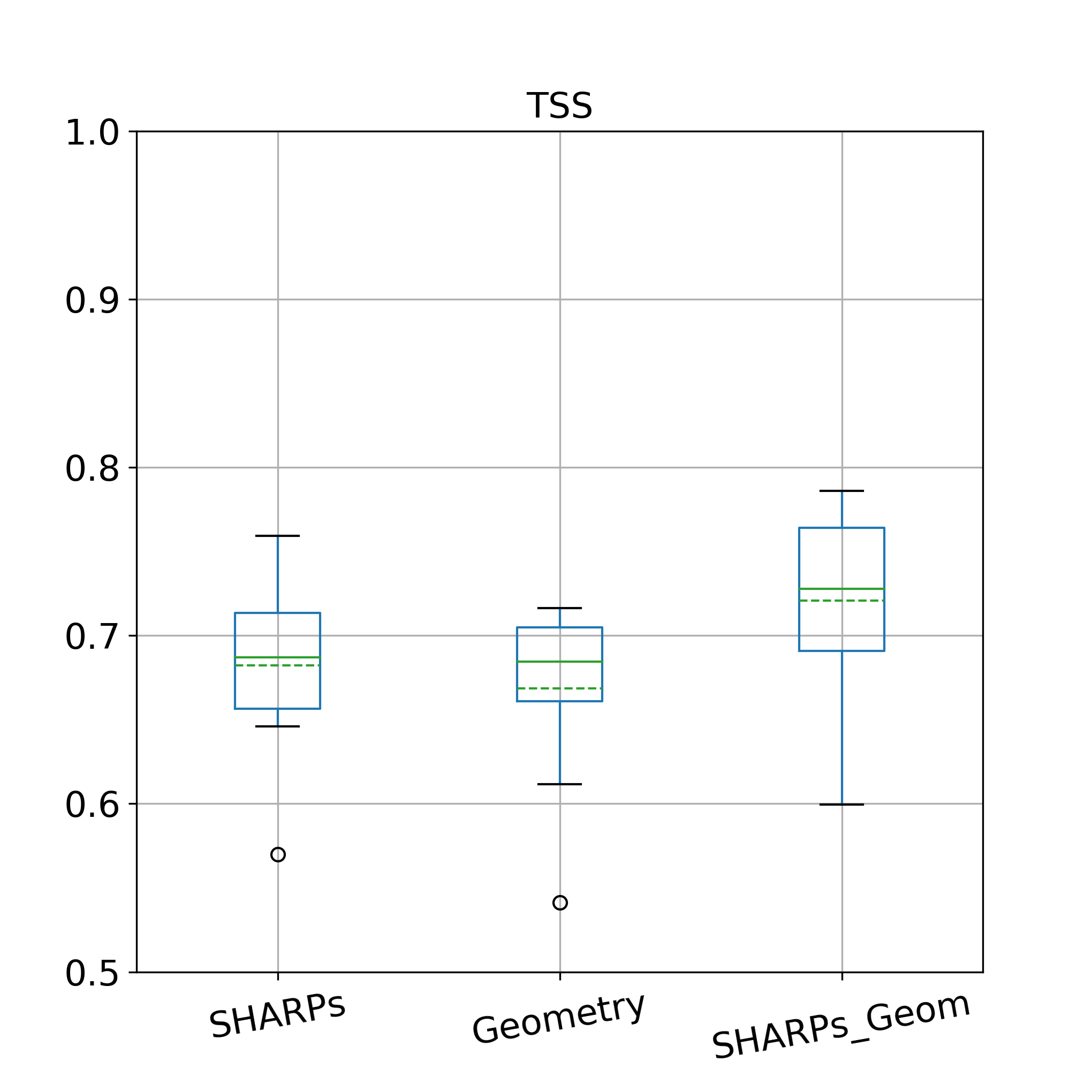}
		}
		\subfloat[Topology Experiments]{
			\includegraphics[width=0.36\textwidth]{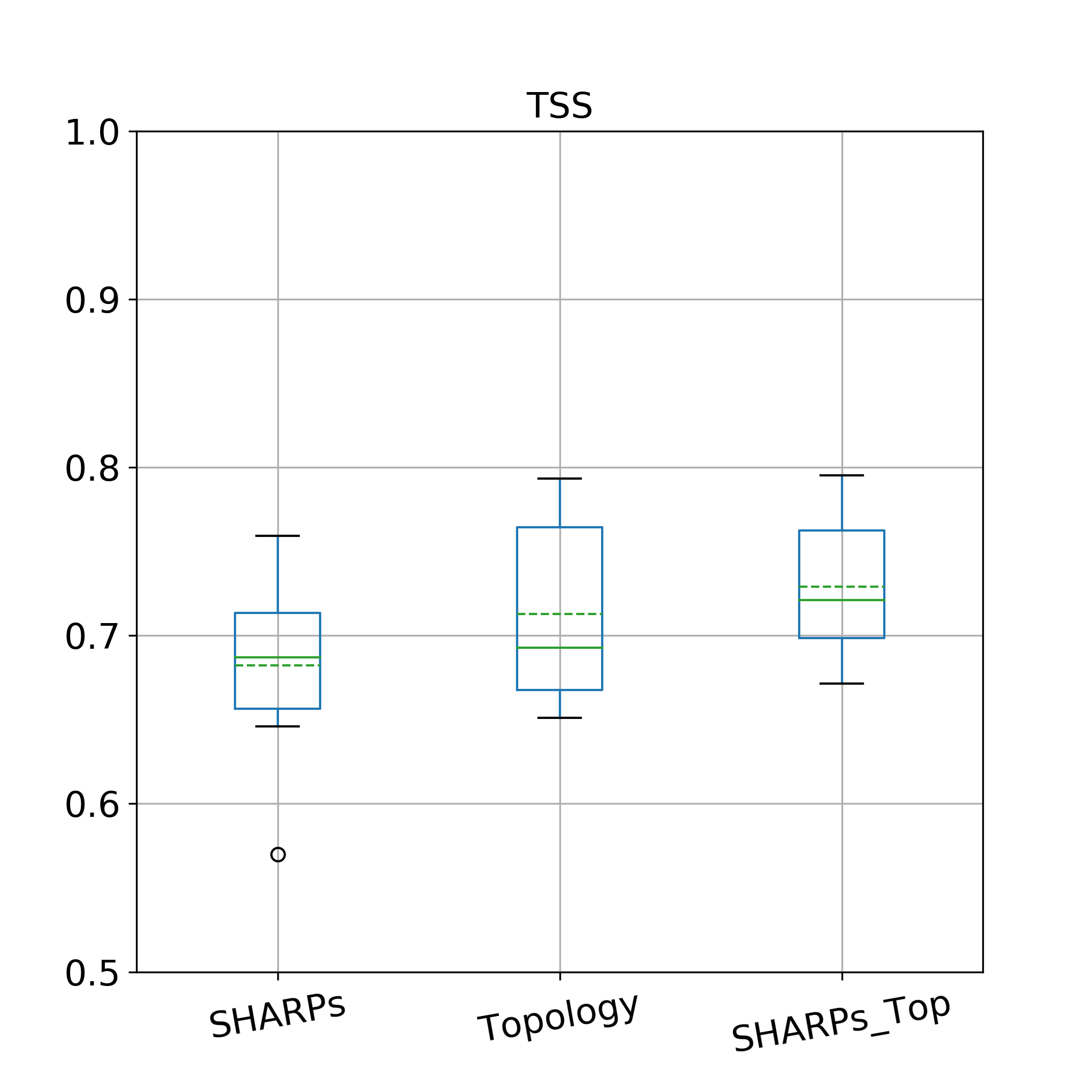}
		}
	\end{center}
	\caption{\revisiona{Box-and-whisker plots of the TSS scores
            for different feature sets used across the geometry and
            topology experiments, reported for ten different
            training/testing splits generated from the same
            dataset. The green solid and dotted lines indicate the
            median and the mean of the TSS scores, respectively, for
            each plot. In plots (a) and (b), the label ``SHARPs\_Geom"
            denotes a combined feature set of SHARPs and geometry-based
            features, while ``SHARPs\_Top" denotes a combined feature set 
            of SHARPs and topology-based features respectively.}}
	\label{fig:tss}
\end{figure}

\begin{figure}
	\begin{center}
		\hspace*{-10mm}
		\subfloat[Geometry Experiments]{
			\includegraphics[width=0.36\textwidth]{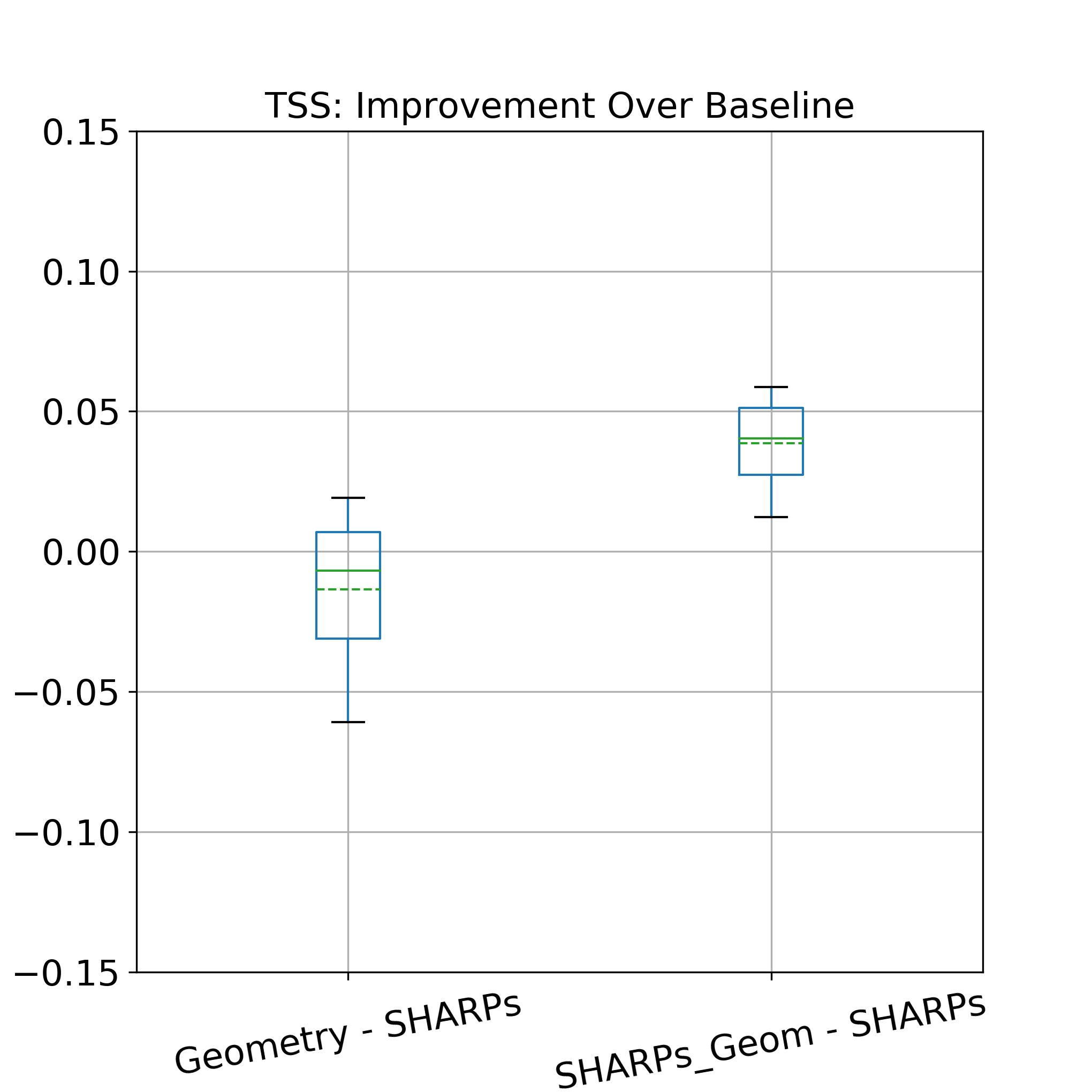}
		}
		\subfloat[Topology Experiments]{
			\includegraphics[width=0.36\textwidth]{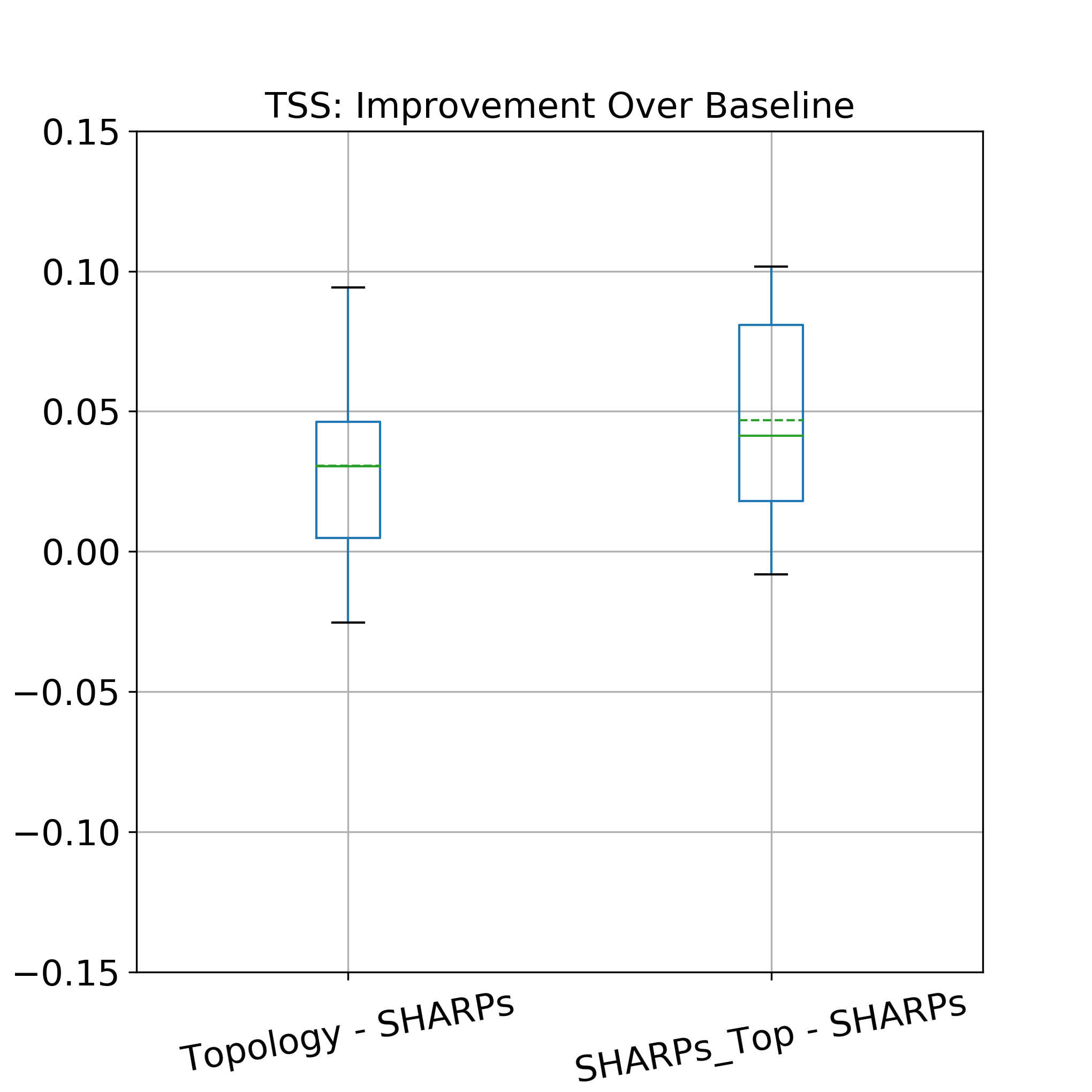}
		}
	\end{center}
	\caption{\revisiona{Box-and-whisker plots of the TSS score
            improvements for the developed feature sets over the
            baseline (SHARPs-only) feature set, reported for ten different
            training/testing splits generated from the same dataset.
            The green solid and dotted lines indicate the median and
            the mean of the TSS score improvements, respectively, for
            each plot. In plots (a) and (b), the label ``SHARPs\_Geom"
            denotes a combined feature set of SHARPs and geometry-based
            features, while ``SHARPs\_Top" denotes a combined feature set 
            of SHARPs and topology-based features respectively.}}
	\label{fig:tss_improvement}
\end{figure}

To study the utility of geometry- and topology-based features in
solar-flare prediction with the ANN described in the previous
sections, we performed a number of experiments.  After 50 training
epochs in each, we computed the True Skill Statistic
\citep{Woodcock1976} (also known as the Hanssen-Kuiper skill score) on
the testing set, which is widely used in solar-flare prediction
studies:
\[{\rm TSS} = \frac{{\rm TP}}{{\rm TP}+{\rm FN}} - \frac{{\rm FP}}{{\rm FP}+{\rm TN}}\]
\noindent where TP and FP (true and false positives) are,
respectively, the images that are classified correctly and
incorrectly as flaring.  Similarly TN and FN are the images
classified correctly and incorrectly as non-flaring.  Note that $-1
\le$ TSS $\le 1$ and TSS $= -1$ only when TP $=$ TN $= 0$, so that every
prediction is wrong; and TSS $= 1$ only when FP $=$ FN $= 0$, so that
every prediction is correct. The skill score reflects an accuracy
relative to a reference forecast that is designed such that both random
forecasts and unskilled forecasts (always predict majority class) have a
score of $0$. When TSS $= 0$ the prediction no
better than chance in the sense that the ``hit rate" is the same as
the ``false alarm rate.''

\revisiona{We performed suites of ten different training/testing
  experiments with different combinations of these three feature sets.
The baseline experiment used the corpus of the 19 SHARPs features of
Tbl.~\ref{tab:sharps-features}.  The mean TSS score for this was
$0.68$ with the range of $[0.57, 0.76]$.  This is in line with other
machine-learning based flare-prediction methods \citep{leka2019a},
indicating that our neural net is a useful test case for this
comparison study.}

\revisiona{Training and validating the same ANN with the 16
  geometry-based features listed in Tbl.~\ref{tab:geom-features}
  yielded a mean TSS value of $0.67$ with a range of $[0.54, 0.72]$,
  suggesting that these geometry-based features, surprisingly, do only
  slightly worse than the SHARPs features, which were constructed by
  solar physics experts to be their best characterizations of an
  active region for the purpose of eruption prediction.  A TSS of
  $0.67$ is also in line with the current {\sl ad hoc} prediction
  methods that employ human experts, qualitative classifications, and
  historical lookup tables \citep{leka2019a}.
  It is encouraging that abstract geometric features, captured
  automatically by an algorithm, allow a machine-learning method to
  match that score.}

\revisiona{To determine how the geometry-based features worked in
  conjunction with the SHARPs features, we repeated these
  experiments using both the SHARPs metadata features and the
  geometry-based features (again on the same data sets), which raised
  the mean TSS score to $0.72$, with a range of $[0.60, 0.79]$.  That
  is, there is a synergy between these two feature sets: the
  combination works better than either one alone, indicating that when
  physical properties of the magnetic field are augmented with
  geometric properties of the active region, the predictive potential
  of a given magnetogram increases.  The box-and-whisker plots in
  panel (a) of Fig.~\ref{fig:tss} provide a graphical comparison of
  the TSS scores reported in this paragraph and the previous one.}

\revisiona{We then trained and tested the ANN using our topology-based
  features, alone and in combination with the SHARPs features.  The
  results are summarized in Fig.~\ref{fig:tss}(b).  Topological
  features alone yielded a mean TSS of $0.72$ with a range of
  $[0.65, 0.79]$.  This is a clear improvement over the performance of
  the ANN with only the baseline SHARPs features.  In combination with
  the SHARPs features, the topological features improved the mean TSS
  score to $0.73$ with a range of $[0.67, 0.80]$.  Comparing the two
  panels of Fig.~\ref{fig:tss}, one can observe that this was also a
  slight improvement over the SHARPs-geometry combination.}

\revisiona{While the above analysis describes the aggregate TSS
  comparison, it is useful to understand how the developed feature sets
  (geometry-based, topology-based and their combination feature set 
  counterparts with SHARPs features) 
  perform with respect to the SHARPs-only baseline in each of the ten 
  training/testing splits. The
  boxplots in Fig.~\ref{fig:tss_improvement} show the improvement in
  TSS scores for the geometry-based, topology-based and the
  combination feature sets with respect to the baseline
  TSS.  This improvement is determined by subtracting the SHARPs-only
  TSS score for each generated training/testing data split from the
  TSS score of the developed feature set using the same
  training/testing split. For the developed feature sets, the mean
  improvement of the TSS over the baseline is in line with the above
  analysis. The geometry-based mean TSS deteriorates by 0.01 while the
  SHARPs-geometry combination show a mean TSS improvement of 0.04. The
  topology-based features and the SHARPs-topology combination show a
  mean TSS improvement of 0.03 and 0.05 respectively.}

The results of these experiments confirm our intuition that abstract
measures of the shapes of magnetogram structures can be indicative of
eruption potential.  This is additionally satisfying since the
historical method of assessing the eruptive potential of an active
region relies on human forecasters essentially analyzing ``complexity
of shape'' in a qualitative manner.  Our results demonstrate that
computational geometry and computational topology have captured this
in a quantitative and repeatable algorithm.  This not only points the
way forward to a more robust set of features for machine-learning base
eruption prediction architectures.  It may even lead to new and
more-effective way to classify sunspot active regions.

\revisiona{In this section, we have relied on TSS as a sole metric for
  comparing the performance of the model for different feature
  sets. An extension of this analysis to other metrics is described in
  Appendix~\ref{sec:appendix_results_all_metrics}.  The scores for
  these alternative metrics, presented in
  Tbl.~\ref{tab:geometry_results_all_metrics} and
  Tbl.~\ref{tab:topology_results_all_metrics}, indicate that their
  trends across the different feature sets are similar to the TSS
  trends discussed above. Additionally, Appendix~\ref{sec:appendix_feature_ranking}
  describes the predictive power of individual features from the 
  different feature sets using the Fisher score --- a univariate feature 
  ranking method. As shown in Figures~\ref{fig:feature_ranking}(a) and 
  Figures~\ref{fig:feature_ranking}(b), 9 of the geometry-based features 
  make it to the top 15 highest ranking features when ranking SHARPs and
  geometry-based features, whereas 6 of the top 15 features belong to the
  topology feature set in the SHARPs-topology ranking experiment.}

\section{Conclusions}
\label{sec:concl}

We have shown that abstract spatial properties of magnetograms can be
useful in machine-learning methods for solar-flare prediction.  We
extract these properties from magnetograms using computational
geometry and computational topology techniques, producing feature
vectors that compete well with the traditional physics-based SHARPs
features for the purposes of machine learning methods for flare
prediction.  
A neural net trained on a large corpus of active regions from HMI,
preprocessed using these techniques, classifies magnetograms as flare
precursors with a slightly higher True Skill Statistic score than the
same model using traditional SHARPs features---the educated assessment
by human experts as to the best information with which to characterize
an active region.  Combining shape- and physics-based features further
improved the TSS scores, indicating a synergy between the two types
of information.  

The power of these highly abstract classifications of structure may be
surprising, particularly in view of the fact that magnetograms are
just the boundary conditions of the full fields whose complicated
dynamics are what produce eruptions. Even so, active region shape has
fundamental, meaningful connections to the physics leading up to an
eruption.  As an active region emerges, becoming progressively larger
and more complex, the shapes of the opposite polarity structures on a
magnetogram capture the evolution of the photospheric field.  Flares
occur when that field forces a rearrangement of the fields in the
corona, where magnetic reconnection occurs. Shape is what human forecasters
use in their classifications, and topology can codify structural complexity
in a formal and yet practical manner that makes it ideally suited for capturing
this richness. In current operational practice, an expert examines sunspot
images and/or magnetograms, classifies  
\revisiona{a sunspot active region} according to taxonomies
developed \revisiona{empirically,} 
and then uses look-up tables of historical
probabilities to say whether or not it will erupt within 
\revisiona{a future time period.}
\revisiona{In a sense, our approach systematizes this ``human-in-the-loop''
forecasting approach by applying the mathematical concepts of topology
to address the shape and connectivity of {\sl sets} derived from the structure of the
photospheric magnetic field.}

In a future paper we plan to validate our results 
\revisiona{using a number of variants of our neural network architecture.}  In
addition, our results can be given a firmer statistical basis by using
more training and testing partitions of the HMI data, different
initial conditions for the network weights, and different numbers of
training epochs.  Finally, neural networks are only one type of
machine-learning model: we will obviously need to compare with other
methods, like SVMs, decision trees, \revisiona{and convolutional
  neural networks (CNNs)}, before making broader claims about the
general utility of shape-based features in machine learning.

There are also a number of data issues that require more attention.
\revisiona{Firstly, in our current approach, we simply discard the
  magnetograms containing invalid pixel data.} Addressing such
pixel-level anomalies in the data---e.g., by smoothing---would be a
good next step here. \revisiona{This will help prevent loss of any potentially
interesting samples and allow us to work with larger datasets.}  It
will also be important to address the appropriate choice of a temporal
resolution: is the selection of every fifth sample (i.e., hourly
resolution) optimal?  This could be carried out in a purely
mathematical fashion, via expert analysis of persistence diagrams, but
it would also be useful to evaluate the effects of different temporal
cadences on the TSS of the neural-net model.

Another important issue is to address is {\sl temporal evolution}: all
current ML-based flare prediction methods, including those reported in
this paper, use only single snapshots of the field.  The dynamics of
the structure of an active region---the progression through time of
the shapes and their relationships---could be captured using
topological approaches that track those structures through time and
space, such as the CROCKER plots of \cite{topaz}.

Last but not least: the geometric and topological feature sets used
here are only a first cut; extending them to capture different aspects
of the shape of an active region could improve the results. 
For example, computational geometry can extract the curvatures of the boundaries of the
different polarity regions.
In terms of topology, the possibilities include higher-dimensional Betti numbers,
alternative complex constructions, and different ways of featurizing persistence diagrams.
These could give us new tools to address the structural complexity of
active regions that are potentially evolving towards an eruption.

\begin{acknowledgements}
This material is based upon work sponsored by the National Science
Foundation Award (Grant No. CMMI 1537460 opinions, findings, and
conclusions or recommendations expressed in this material are those of
the author(s) and do not necessarily reflect the views of the NSF.
The University of Colorado Space Weather Technology, Research,
and Education Center (SWx TREC) is supported by a Grand Challenge grant 
from the Chancellor of the University of Colorado and hosted in the College
of Engineering and Applied Sciences (CEAS) on the Boulder campus. 
\end{acknowledgements}

\bibliography{references,Berger_refs}

\begin{thebibliography}{79}
\providecommand{\natexlab}[1]{#1}
\providecommand{\url}[1]{\texttt{#1}}
\providecommand{\urlprefix}{URL }
\providecommand{\eprint}[2][]{\url{#2}}

\bibitem[{Adams et~al.(2017)Adams, Emerson, Kirby, Neville, Peterson, Shipman,
  Chepushtanova, Hanson, Motta, and Ziegelmeier}]{Adams2017}
Adams, H., T.~Emerson, M.~Kirby, R.~Neville, C.~Peterson, P.~Shipman,
  S.~Chepushtanova, E.~Hanson, F.~Motta, and L.~Ziegelmeier, 2017.
\newblock Persistence Images: A Stable Vector Representation of Persistent
  Homology.
\newblock \emph{J. Mach. Learn. Res.}, \textbf{18}(1), 218–252.
\newblock 10.5555/3122009.3122017.

\bibitem[{Ahmed et~al.(2010)Ahmed, Qahwaji, Colak, Dudok De~Wit, and
  Ipson}]{Ahmed2010}
Ahmed, O.~W., R.~Qahwaji, T.~Colak, T.~Dudok De~Wit, and S.~Ipson, 2010.
\newblock A new technique for the calculation and 3D visualisation of magnetic
  complexities on solar satellite images.
\newblock \emph{The Visual Computer}, \textbf{26}(5), 385--395.
\newblock 10.1007/s00371-010-0418-1.

\bibitem[{Aulanier et~al.(2005)Aulanier, Pariat, and
  D{\'e}moulin}]{Aulanier:2005}
Aulanier, G., E.~Pariat, and P.~D{\'e}moulin, 2005.
\newblock {Current Sheet Formation in Quasi-separatrix Layers and Hyperbolic
  Flux Tubes}.
\newblock \emph{Astronomy {\&} Astrophysics}, \textbf{444}, 961--976.
\newblock 10.1051/0004-6361:20053600.

\bibitem[{Barnes and Leka(2006)}]{Barnes:2006wo}
Barnes, G., and K.~D. Leka, 2006.
\newblock {Photospheric Magnetic Field Properties of Flaring vs. Non-Flare
  Quiet Active Regions. III. Magnetic Charge Topology Models}.
\newblock \emph{Astrophysical Journal}, \textbf{646}, 1303--1318.
\newblock 10.1086/510282.

\bibitem[{Barnes et~al.(2016)Barnes, Leka, Schrijver, Colak, Qahwaji
  et~al.}]{Barnes:2016bu}
Barnes, G., K.~D. Leka, C.~J. Schrijver, T.~Colak, R.~Qahwaji, et~al., 2016.
\newblock {A Comparison of Flare Forecasting Methods. I. Results from the
  {\textquotedblleft}All-Clear{\textquotedblright} Workshop}.
\newblock \emph{Astrophysical Journal}, \textbf{829}(2), 1--32.
\newblock 10.3847/0004-637x/829/2/89.

\bibitem[{Barnes et~al.(2005)Barnes, Longcope, and Leka}]{Barnes:2005wv}
Barnes, G., D.~W. Longcope, and K.~D. Leka, 2005.
\newblock {Implementing a Magnetic Charge Topology Model for Solar Active
  Regions}.
\newblock \emph{Astrophysical Journal}, \textbf{629}, 561--571.
\newblock 10.1086/431175.

\bibitem[{Benvenuto et~al.(2018)Benvenuto, Piana, Campi, and
  Massone}]{Benvenuto2018}
Benvenuto, F., M.~Piana, C.~Campi, and A.~M. Massone, 2018.
\newblock A Hybrid Supervised/Unsupervised Machine Learning Approach to Solar
  Flare Prediction.
\newblock \emph{Astrophysical Journal}, \textbf{853}(1), 90.
\newblock 10.3847/1538-4357/aaa23c.

\bibitem[{Bhavsar and Lauer(1996)}]{Bhavsar1996}
Bhavsar, S.~P., and D.~A. Lauer.
\newblock Analysis of the CFA ``Great Wall'' Using The Minimal Spanning Tree,
  517--519.
\newblock Springer Netherlands, Dordrecht, 1996.
\newblock ISBN 978-94-009-0145-2.
\newblock \url{10.1007/978-94-009-0145-2_66}.

\bibitem[{Bobra and Couvidat(2015)}]{Bobra:2015fn}
Bobra, M.~G., and S.~Couvidat, 2015.
\newblock {Solar Flare Prediction Using SDO/HMI Vector Magnetic Field Data with
  a Machine-Learning Algorithm}.
\newblock \emph{Astrophysical Journal}, \textbf{798}(2), 135--11.
\newblock 10.1088/0004-637x/798/2/135.

\bibitem[{Bobra et~al.(2014)Bobra, Sun, Hoeksema, Turmon, Liu, Hayashi, Barnes,
  and Leka}]{Bobra:2014dn}
Bobra, M.~G., X.~Sun, J.~T. Hoeksema, M.~Turmon, Y.~Liu, K.~Hayashi, G.~Barnes,
  and K.~D. Leka, 2014.
\newblock {The Helioseismic and Magnetic Imager (HMI) Vector Magnetic Field
  Pipeline: SHARPs {\textendash} Space-Weather HMI Active Region Patches}.
\newblock \emph{Solar Physics}, \textbf{289}(9), 3549--3578.
\newblock 10.1007/s11207-014-0529-3.

\bibitem[{{Boucheron} et~al.(2015){Boucheron}, {Al-Ghraibah}, and
  {McAteer}}]{boucheron2015}
{Boucheron}, L.~E., A.~{Al-Ghraibah}, and R.~T.~J. {McAteer}, 2015.
\newblock {Prediction of Solar Flare Size and Time-to-Flare Using Support
  Vector Machine Regression}.
\newblock \emph{Astrophysical Journal}, \textbf{812}, 51.
\newblock 10.1088/0004-637X/812/1/51.

\bibitem[{Bubenik(2015)}]{Bubenik2015}
Bubenik, P., 2015.
\newblock Statistical Topological Data Analysis Using Persistence Landscapes.
\newblock \emph{J. Mach. Learn. Res.}, \textbf{16}(1), 77–102.
\newblock 10.5555/2789272.2789275.

\bibitem[{Campi et~al.(2019)Campi, Benvenuto, Massone, Bloomfield, Georgoulis,
  and Piana}]{Campi2019}
Campi, C., F.~Benvenuto, A.~M. Massone, D.~S. Bloomfield, M.~K. Georgoulis, and
  M.~Piana, 2019.
\newblock Feature Ranking of Active Region Source Properties in Solar Flare
  Forecasting and the Uncompromised Stochasticity of Flare Occurrence.
\newblock \emph{Astrophysical Journal}, \textbf{883}(2), 150.
\newblock 10.3847/1538-4357/ab3c26.

\bibitem[{Camporeale(2019)}]{Camporeale2019}
Camporeale, E., 2019.
\newblock {The Challenge of Machine Learning in Space Weather: Nowcasting and
  Forecasting}.
\newblock \emph{Space Weather}, \textbf{17}(8), 1166--1207.
\newblock 10.1029/2018SW002061.

\bibitem[{{Carri{\`e}re} et~al.(2019){Carri{\`e}re}, {Chazal}, {Ike},
  {Lacombe}, {Royer}, and {Umeda}}]{Carriere2019}
{Carri{\`e}re}, M., F.~{Chazal}, Y.~{Ike}, T.~{Lacombe}, M.~{Royer}, and
  Y.~{Umeda}, 2019.
\newblock {PersLay: A Simple and Versatile Neural Network Layer for Persistence
  Diagrams}.
\newblock \emph{arXiv e-prints}, arXiv:1904.09378.

\bibitem[{Carri\`{e}re et~al.(2017)Carri\`{e}re, Cuturi, and
  Oudot}]{Carriere2017}
Carri\`{e}re, M., M.~Cuturi, and S.~Oudot, 2017.
\newblock Sliced Wasserstein Kernel for Persistence Diagrams.
\newblock In Proceedings of the 34th International Conference on Machine
  Learning - Volume 70, ICML’17, 664–673. JMLR.org.
\newblock 10.5555/3305381.3305450.

\bibitem[{Chazal et~al.(2014)Chazal, Fasy, Lecci, Rinaldo, and
  Wasserman}]{Chazal2014}
Chazal, F., B.~T. Fasy, F.~Lecci, A.~Rinaldo, and L.~Wasserman, 2014.
\newblock Stochastic Convergence of Persistence Landscapes and Silhouettes.
\newblock In Proceedings of the Thirtieth Annual Symposium on Computational
  Geometry, SOCG'14, 474:474--474:483. ACM, New York, NY, USA.
\newblock ISBN 978-1-4503-2594-3.
\newblock 10.1145/2582112.2582128.

\bibitem[{Chollet et~al.(2015)}]{Chollet2015}
Chollet, F., et~al., 2015.
\newblock Keras.
\newblock \url{https://github.com/fchollet/keras}.

\bibitem[{Colak and Qahwaji(2009)}]{colak2009}
Colak, T., and R.~Qahwaji, 2009.
\newblock Automated Solar Activity Prediction: A hybrid computer platform using
  machine learning and solar imaging for automated prediction of solar flares.
\newblock \emph{Space Weather}, \textbf{7}(6).
\newblock 10.1029/2008SW000401.

\bibitem[{{Crown}(2012)}]{Crown2012}
{Crown}, M.~D., 2012.
\newblock {Validation of the NOAA Space Weather Prediction Center's solar flare
  forecasting look-up table and forecaster-issued probabilities}.
\newblock \emph{Space Weather}, \textbf{10}, S06006.
\newblock 10.1029/2011SW000760.

\bibitem[{de~Silva and Ghrist(2007)}]{deSilva07}
de~Silva, V., and R.~Ghrist, 2007.
\newblock Coverage in sensor networks via persistent homology.
\newblock \emph{Algebr. Geom. Topol.}, \textbf{7}(1), 339--358.
\newblock 10.2140/agt.2007.7.339.

\bibitem[{{Demoulin} et~al.(1996){Demoulin}, {Henoux}, {Priest}, and
  {Mandrini}}]{Demoulin:1996}
{Demoulin}, P., J.~C. {Henoux}, E.~R. {Priest}, and C.~H. {Mandrini}, 1996.
\newblock {Quasi-Separatrix Layers in Solar Fares. I. Method.}
\newblock \emph{Astronomy {\&} Astrophysics}, \textbf{308}, 643--655.

\bibitem[{DeRosa et~al.(2009)DeRosa, Schrijver, Barnes, Leka, Lites
  et~al.}]{DeRosa:2009ce}
DeRosa, M.~L., C.~J. Schrijver, G.~Barnes, K.~D. Leka, B.~W. Lites, et~al.,
  2009.
\newblock {A Critical Assessment of Nonlinear Force-Free Field Modeling of the
  Solar Corona for Active Region 10953}.
\newblock \emph{Astrophysical Journal}, \textbf{696}(2), 1780--1791.
\newblock 10.1088/0004-637x/696/2/1780.

\bibitem[{Devogele et~al.(2015)Devogele, P.~Rivet, Tanga, Bendjoya, Surdej,
  Bartczak, and Hanuš}]{Devogele}
Devogele, M., J.~P.~Rivet, P.~Tanga, P.~Bendjoya, J.~Surdej, P.~Bartczak, and
  J.~Hanuš, 2015.
\newblock {A Method to Search for Large-Scale Concavities in Asteroid Shape
  Models}.
\newblock \emph{Monthly Notices of the Royal Astronomical Society},
  \textbf{453}.
\newblock 10.1093/mnras/stv1740.

\bibitem[{Duchi et~al.(2011)Duchi, Hazan, and Singer}]{Duchi2011}
Duchi, J., E.~Hazan, and Y.~Singer, 2011.
\newblock Adaptive Subgradient Methods for Online Learning and Stochastic
  Optimization.
\newblock \emph{J. Mach. Learn. Res.}, \textbf{12}(null), 2121–2159.
\newblock 10.5555/1953048.2021068.

\bibitem[{Edelsbrunner et~al.(2000)Edelsbrunner, Letscher, and
  Zomorodian}]{Edelsbrunner00}
Edelsbrunner, H., D.~Letscher, and A.~Zomorodian, 2000.
\newblock {Topological Persistence and Simplification}.
\newblock \emph{Discrete \& Computational Geometry}, \textbf{28}, 511--533.
\newblock 10.1007/s00454-002-2885-2.

\bibitem[{{Florios} et~al.(2018){Florios}, {Kontogiannis}, {Park}, {Guerra},
  {Benvenuto}, {Bloomfield}, and {Georgoulis}}]{Florios2018}
{Florios}, K., I.~{Kontogiannis}, S.-H. {Park}, J.~A. {Guerra}, F.~{Benvenuto},
  D.~S. {Bloomfield}, and M.~K. {Georgoulis}, 2018.
\newblock {Forecasting Solar Flares Using Magnetogram-based Predictors and
  Machine Learning}.
\newblock \emph{\solphys}, \textbf{293}(2), 28.
\newblock 10.1007/s11207-018-1250-4.

\bibitem[{Forrest(1971)}]{Forrest187}
Forrest, A.~R., 1971.
\newblock {II. Current Developments in the Design and Production of
  Three-Dimensional Curved Objects --- Computational Geometry}.
\newblock \emph{Proceedings of the Royal Society of London A: Mathematical,
  Physical and Engineering Sciences}, \textbf{321}(1545), 187--195.
\newblock 10.1098/rspa.1971.0025.

\bibitem[{Gallagher et~al.(2002)Gallagher, Moon, and Wang}]{gallagher2002}
Gallagher, P., Y.-J. Moon, and H.~Wang, 2002.
\newblock Active-Region Monitoring and Flare Forecasting - I. Data Processing
  and First Results.
\newblock \emph{Sol. Phys.}, \textbf{209}, 171--183.
\newblock 10.1023/A:1020950221179.

\bibitem[{Ghrist(2008)}]{Ghrist08}
Ghrist, R., 2008.
\newblock {Barcodes: The Persistent Topology of Data}.
\newblock \emph{Bulletin of the American Mathematical Society}, \textbf{45}(1),
  61--75.
\newblock 10.1090/S0273-0979-07-01191-3.

\bibitem[{Glorot and Bengio(2010)}]{Glorot2010}
Glorot, X., and Y.~Bengio, 2010.
\newblock Understanding the difficulty of training deep feedforward neural
  networks.
\newblock In Y.~W. Teh and M.~Titterington, eds., Proceedings of the Thirteenth
  International Conference on Artificial Intelligence and Statistics, vol.~9 of
  \emph{Proceedings of Machine Learning Research}, 249--256. PMLR, Chia Laguna
  Resort, Sardinia, Italy.
\newblock 10.1.1.207.2059.

\bibitem[{Gu et~al.(2011)Gu, Li, and Han}]{Gu2012}
Gu, Q., Z.~Li, and J.~Han, 2011.
\newblock Generalized Fisher Score for Feature Selection.
\newblock In Proceedings of the Twenty-Seventh Conference on Uncertainty in
  Artificial Intelligence, UAI’11, 266–273. AUAI Press, Arlington,
  Virginia, USA.
\newblock ISBN 9780974903972.
\newblock 10.5555/3020548.3020580.

\bibitem[{Guerra et~al.(2018)Guerra, Park, Gallagher, Kontogiannis, Georgoulis,
  and Bloomfield}]{Guerra2018}
Guerra, J.~A., S.-H. Park, P.~T. Gallagher, I.~Kontogiannis, M.~K. Georgoulis,
  and D.~S. Bloomfield, 2018.
\newblock Active Region Photospheric Magnetic Properties Derived from
  Line-of-Sight and Radial Fields.
\newblock \emph{Solar Physics}, \textbf{293}(1), 9.
\newblock 10.1007/s11207-017-1231-z.

\bibitem[{{Hale} et~al.(1919){Hale}, {Ellerman}, {Nicholson}, and
  {Joy}}]{Hale:1919}
{Hale}, G.~E., F.~{Ellerman}, S.~B. {Nicholson}, and A.~H. {Joy}, 1919.
\newblock {The Magnetic Polarity of Sun-Spots}.
\newblock \emph{\apj}, \textbf{49}, 153.
\newblock 10.1086/142452.

\bibitem[{Haykin(1998)}]{Haykin1998}
Haykin, S., 1998.
\newblock Neural Networks: A Comprehensive Foundation.
\newblock Prentice Hall PTR, USA, 2nd edn.
\newblock ISBN 0132733501.
\newblock 10.5555/521706.

\bibitem[{Huang et~al.(2018)Huang, Wang, Xu, Liu, Li, and Dai}]{Huang2018}
Huang, X., H.~Wang, L.~Xu, J.~Liu, R.~Li, and X.~Dai, 2018.
\newblock Deep Learning Based Solar Flare Forecasting Model. I. Results for
  Line-of-sight Magnetograms.
\newblock \emph{Astrophysical Journal}, \textbf{856}(1), 7.
\newblock 10.3847/1538-4357/aaae00.

\bibitem[{{Jonas} et~al.(2018){Jonas}, {Bobra}, {Shankar}, {Todd Hoeksema}, and
  {Recht}}]{Jonas:2018}
{Jonas}, E., M.~{Bobra}, V.~{Shankar}, J.~{Todd Hoeksema}, and B.~{Recht},
  2018.
\newblock {Flare Prediction Using Photospheric and Coronal Image Data}.
\newblock \emph{\solphys}, \textbf{293}(3), 48.
\newblock 10.1007/s11207-018-1258-9.

\bibitem[{Kaczynski et~al.(2004)Kaczynski, Mischaikow, and
  Mrozek}]{Kaczynski04}
Kaczynski, T., K.~Mischaikow, and M.~Mrozek, 2004.
\newblock Computational Homology, vol. 157 of \emph{Applied Mathematical
  Sciences}.
\newblock Springer-Verlag, New York.
\newblock 10.1007/B97315.

\bibitem[{{Kingma} and {Ba}(2014)}]{Kingma2014}
{Kingma}, D.~P., and J.~{Ba}, 2014.
\newblock {Adam: A Method for Stochastic Optimization}.
\newblock \emph{arXiv e-prints}, arXiv:1412.6980.

\bibitem[{Knyazeva et~al.(2011)Knyazeva, Makarenko, and Livshits}]{Knyazeva11}
Knyazeva, I.~S., N.~G. Makarenko, and M.~A. Livshits, 2011.
\newblock {Detection of New Emerging Magnetic Flux from the Topology of
  {SOHO/MDI} Magnetograms}.
\newblock \emph{Astronomy Reports}, \textbf{55}(5), 463.
\newblock 10.1134/S1063772911050040.

\bibitem[{Kontogiannis et~al.(2018)Kontogiannis, Georgoulis, Park, and
  Guerra}]{Kontogiannis2018}
Kontogiannis, I., M.~K. Georgoulis, S.-H. Park, and J.~A. Guerra, 2018.
\newblock Testing and Improving a Set of Morphological Predictors of Flaring
  Activity.
\newblock \emph{Solar Physics}, \textbf{293}(6), 96.
\newblock 10.1007/s11207-018-1317-2.

\bibitem[{Kusano et~al.(2016)Kusano, Fukumizu, and Hiraoka}]{Kusano2016}
Kusano, G., K.~Fukumizu, and Y.~Hiraoka, 2016.
\newblock Persistence Weighted Gaussian Kernel for Topological Data Analysis.
\newblock In Proceedings of the 33rd International Conference on International
  Conference on Machine Learning - Volume 48, ICML’16, 2004–2013. JMLR.org.
\newblock 10.5555/3045390.3045602.

\bibitem[{Le and Yamada(2018)}]{Le2018}
Le, T., and M.~Yamada, 2018.
\newblock Persistence Fisher Kernel: A Riemannian Manifold Kernel for
  Persistence Diagrams.
\newblock In Proceedings of the 32nd International Conference on Neural
  Information Processing Systems, NIPS’18, 10028–10039. Curran Associates
  Inc., Red Hook, NY, USA.
\newblock {10.5555/3327546.3327666}.

\bibitem[{{Leka} et~al.(2019{\natexlab{a}}){Leka}, {Park}, {Kusano}, {Andries},
  {Barnes} et~al.}]{leka2019a}
{Leka}, K.~D., S.-H. {Park}, K.~{Kusano}, J.~{Andries}, G.~{Barnes}, et~al.,
  2019{\natexlab{a}}.
\newblock {A Comparison of Flare Forecasting Methods. II. Benchmarks, Metrics,
  and Performance Results for Operational Solar Flare Forecasting Systems}.
\newblock \emph{\apjs}, \textbf{243}(2), 36.
\newblock 10.3847/1538-4365/ab2e12.

\bibitem[{{Leka} et~al.(2019{\natexlab{b}}){Leka}, {Park}, {Kusano}, {Andries},
  {Barnes} et~al.}]{leka2019b}
{Leka}, K.~D., S.-H. {Park}, K.~{Kusano}, J.~{Andries}, G.~{Barnes}, et~al.,
  2019{\natexlab{b}}.
\newblock {A Comparison of Flare Forecasting Methods. III. Systematic Behaviors
  of Operational Solar Flare Forecasting Systems}.
\newblock \emph{\apj}, \textbf{881}(2), 101.
\newblock 10.3847/1538-4357/ab2e11.

\bibitem[{Longcope(2005)}]{Longcope05}
Longcope, D.~W., 2005.
\newblock Topological Methods for the Analysis of Solar Magnetic Fields.
\newblock \emph{Living Reviews in Solar Physics}, \textbf{2}(1), 7.
\newblock 10.12942/lrsp-2005-7.

\bibitem[{Makarenko et~al.(2014)Makarenko, Malkova, Machin, Knyazeva, and
  Makarenko}]{Makarenko14}
Makarenko, N., D.~Malkova, M.~Machin, I.~Knyazeva, and I.~Makarenko, 2014.
\newblock {Methods of Computational Topology for the Analysis of Dynamics of
  Active Regions of the Sun}.
\newblock \emph{Journal of Mathematical Sciences}, \textbf{203}(6), 806--815.
\newblock 10.1007/s10958-014-2170-y.

\bibitem[{{McAteer} et~al.(2010){McAteer}, {Gallagher}, and
  {Conlon}}]{mcateer2010}
{McAteer}, R.~T.~J., P.~T. {Gallagher}, and P.~A. {Conlon}, 2010.
\newblock {Turbulence, complexity, and solar flares}.
\newblock \emph{Advances in Space Research}, \textbf{45}, 1067--1074.
\newblock 10.1016/j.asr.2009.08.026.

\bibitem[{{McIntosh}(1990)}]{McIntosh:1990wu}
{McIntosh}, P.~S., 1990.
\newblock {The Classification of Sunspot Groups}.
\newblock \emph{Solar Physics}, \textbf{125}, 251--267.
\newblock 10.1007/BF00158405.

\bibitem[{Metcalf et~al.(2008)Metcalf, DeRosa, Schrijver, Barnes, van
  Ballegooijen, Wiegelmann, Wheatland, Valori, , and
  McTiernan}]{Metcalf:2008df}
Metcalf, T.~R., M.~L. DeRosa, C.~J. Schrijver, G.~Barnes, A.~A. van
  Ballegooijen, T.~Wiegelmann, M.~S. Wheatland, G.~Valori, , and J.~M.
  McTiernan, 2008.
\newblock {Nonlinear Force-Free Modeling of Coronal Magnetic Fields. II.
  Modeling a Filament Arcade and Simulated Chromospheric and Photospheric
  Vector Fields}.
\newblock \emph{Solar Physics}, \textbf{247}(2), 269--299.
\newblock 10.1007/s11207-007-9110-7.

\bibitem[{{Nishizuka} et~al.(2018){Nishizuka}, {Sugiura}, {Kubo}, {Den}, and
  {Ishii}}]{Nishizuka:2018}
{Nishizuka}, N., K.~{Sugiura}, Y.~{Kubo}, M.~{Den}, and M.~{Ishii}, 2018.
\newblock {Deep Flare Net (DeFN) Model for Solar Flare Prediction}.
\newblock \emph{Astrophysical Journal}, \textbf{858}, 113.
\newblock 10.3847/1538-4357/aab9a7.

\bibitem[{{Nishizuka} et~al.(2017){Nishizuka}, {Sugiura}, {Kubo}, {Den},
  {Watari}, and {Ishii}}]{Nishizuka:2017}
{Nishizuka}, N., K.~{Sugiura}, Y.~{Kubo}, M.~{Den}, S.~{Watari}, and
  M.~{Ishii}, 2017.
\newblock {Solar Flare Prediction Model with Three Machine-learning Algorithms
  using Ultraviolet Brightening and Vector Magnetograms}.
\newblock \emph{Astrophysical Journal}, \textbf{835}, 156.
\newblock 10.3847/1538-4357/835/2/156.

\bibitem[{Park et~al.(2018)Park, Moon, Shin, Yi, Lim, Lee, and Shin}]{Park2018}
Park, E., Y.-J. Moon, S.~Shin, K.~Yi, D.~Lim, H.~Lee, and G.~Shin, 2018.
\newblock Application of the Deep Convolutional Neural Network to the Forecast
  of Solar Flare Occurrence Using Full-disk Solar Magnetograms.
\newblock \emph{Astrophysical Journal}, \textbf{869}(2), 91.
\newblock 10.3847/1538-4357/aaed40.

\bibitem[{Pedregosa et~al.(2011)Pedregosa, Varoquaux, Gramfort, Michel, Thirion
  et~al.}]{Pedregosa2011}
Pedregosa, F., G.~Varoquaux, A.~Gramfort, V.~Michel, B.~Thirion, et~al., 2011.
\newblock Scikit-Learn: Machine Learning in Python.
\newblock \emph{J. Mach. Learn. Res.}, \textbf{12}(null), 2825–2830.
\newblock 10.5555/1953048.2078195.

\bibitem[{Pesnell et~al.(2011)Pesnell, Thompson, and
  Chamberlin}]{Pesnell:2011ik}
Pesnell, W.~D., B.~J. Thompson, and P.~C. Chamberlin, 2011.
\newblock {The Solar Dynamics Observatory (SDO)}.
\newblock \emph{Solar Physics}, \textbf{275}(1-2), 3--15.
\newblock 10.1007/s11207-011-9841-3.

\bibitem[{Preparata and Shamos(1985)}]{preparata}
Preparata, F.~P., and M.~I. Shamos, 1985.
\newblock Computational Geometry: An Introduction.
\newblock Springer-Verlag, New York.
\newblock 10.1007/978-1-4612-1098-6.

\bibitem[{{Priest} and {D{\'e}moulin}(1995)}]{Priest-Demoulin:1995}
{Priest}, E.~R., and P.~{D{\'e}moulin}, 1995.
\newblock {Three-dimensional Magnetic Reconnection without Null Points. 1.
  Basic Theory of Magnetic Flipping}.
\newblock \emph{Journal of Geophysical Research}, \textbf{100}, 23,443--23,464.
\newblock 10.1029/95JA02740.

\bibitem[{{Reininghaus} et~al.(2015){Reininghaus}, {Huber}, {Bauer}, and
  {Kwitt}}]{Reininghaus2015}
{Reininghaus}, J., S.~{Huber}, U.~{Bauer}, and R.~{Kwitt}, 2015.
\newblock A stable multi-scale kernel for topological machine learning.
\newblock In 2015 IEEE Conference on Computer Vision and Pattern Recognition
  (CVPR), 4741--4748.
\newblock 10.1109/CVPR.2015.7299106.

\bibitem[{Robins et~al.(1998)Robins, Meiss, and Bradley}]{Robins98}
Robins, V., J.~Meiss, and E.~Bradley, 1998.
\newblock {Computing Connectedness: An Exercise in Computational Topology}.
\newblock \emph{Nonlinearity}, \textbf{11}(4), 913--922.

\bibitem[{Robins et~al.(2000)Robins, Meiss, and Bradley}]{Robins00c}
Robins, V., J.~Meiss, and E.~Bradley, 2000.
\newblock {Computing connectedness: Disconnectedness and Discreteness}.
\newblock \emph{Physica D}, \textbf{139}(3-4), 276--300.

\bibitem[{Scherrer et~al.(2011)Scherrer, Schou, Bush, Kosovichev, Bogart
  et~al.}]{Scherrer:2011ji}
Scherrer, P.~H., J.~Schou, R.~I. Bush, A.~G. Kosovichev, R.~S. Bogart, et~al.,
  2011.
\newblock {The Helioseismic and Magnetic Imager (HMI) Investigation for the
  Solar Dynamics Observatory (SDO)}.
\newblock \emph{Solar Physics}, \textbf{275}(1-2), 207--227.
\newblock 10.1007/s11207-011-9834-2.

\bibitem[{Schrijver(2007)}]{Schrijver}
Schrijver, C.~J., 2007.
\newblock A Characteristic Magnetic Field Pattern Associated with All Major
  Solar Flares and Its Use in Flare Forecasting.
\newblock \emph{Astrophysical Journal Letters}, \textbf{655}(2), L117.
\newblock 10.1086/511857.

\bibitem[{Schrijver(2016)}]{Schrijver:2016cs}
Schrijver, C.~J., 2016.
\newblock {The Nonpotentiality of Coronae of Solar Active Regions, the Dynamics
  of the Surface Magnetic Field, and the Potential for Large Flares}.
\newblock \emph{Astrophysical Journal}, \textbf{820}(2), 1--17.
\newblock 10.3847/0004-637x/820/2/103.

\bibitem[{Schrijver et~al.(2008)Schrijver, DeRosa, Metcalf, Barnes, Lites
  et~al.}]{Schrijver:2008tg}
Schrijver, C.~J., M.~L. DeRosa, T.~Metcalf, G.~Barnes, B.~Lites, et~al., 2008.
\newblock {Nonlinear Force-Free Field Modeling of a Solar Active Region around
  the time of a Major Flare and Coronal Mass Ejection}.
\newblock \emph{Astrophysical Journal}, \textbf{675}, 1637--1644.
\newblock 10.1086/527413.

\bibitem[{Schrijver et~al.(2006)Schrijver, DeRosa, Metcalf, Liu, {McTiernan},
  R{\'e}gnier, Valori, Wheatland, and Wiegelmann}]{Schrijver:2006jw}
Schrijver, C.~J., M.~L. DeRosa, T.~R. Metcalf, Y.~Liu, J.~{McTiernan},
  S.~R{\'e}gnier, G.~Valori, M.~S. Wheatland, and T.~Wiegelmann, 2006.
\newblock {Nonlinear Force-Free Modeling of Coronal Magnetic Fields Part I: A
  Quantitative Comparison of Methods}.
\newblock \emph{Solar Physics}, \textbf{235}(1-2), 161--190.
\newblock 10.1007/s11207-006-0068-7.

\bibitem[{Singh et~al.(2008)Singh, Memoli, Ishkhanov, Sapiro, Carlsson, and
  Ringach}]{Singh08}
Singh, G., F.~Memoli, T.~Ishkhanov, G.~Sapiro, G.~Carlsson, and D.~Ringach,
  2008.
\newblock {Topological Analysis of Population Activity in Visual Cortex}.
\newblock \emph{Journal of Vision}, \textbf{8}(8:(11)), 1--18.
\newblock 10.1167/8.8.11.

\bibitem[{Tarr and Longcope(2012)}]{Tarr:2012}
Tarr, L., and D.~Longcope, 2012.
\newblock Calculating Energy Storage Due to Topological Changes in Emerging
  Active Region {NOAA} {AR} 11112.
\newblock \emph{Astrophysical Journal}, \textbf{749}(1), 64.
\newblock 10.1088/0004-637x/749/1/64.

\bibitem[{{Tarr} et~al.(2013){Tarr}, {Longcope}, and {Millhouse}}]{Tarr:2013}
{Tarr}, L., D.~{Longcope}, and M.~{Millhouse}, 2013.
\newblock {Calculating Separate Magnetic Free Energy Estimates for Active
  Regions Producing Multiple Flares: NOAA AR11158}.
\newblock \emph{\apj}, \textbf{770}(1), 4.
\newblock 10.1088/0004-637X/770/1/4.

\bibitem[{Topaz et~al.(2015)Topaz, Ziegelmeier, and Halverson}]{topaz}
Topaz, C.~M., L.~Ziegelmeier, and T.~Halverson, 2015.
\newblock Topological Data Analysis of Biological Aggregation Models.
\newblock \emph{PLOS ONE}, \textbf{10}(5), 1--26.
\newblock 10.1371/journal.pone.0126383.

\bibitem[{{Wang} and {Sheeley Jr.}(1992)}]{Wang:1992}
{Wang}, Y.~M., and N.~R. {Sheeley Jr.}, 1992.
\newblock {On Potential Field Models of the Solar Corona}.
\newblock \emph{Astrophysical Journal}, \textbf{392}, 310.
\newblock 10.1086/171430.

\bibitem[{{Wheatland}(2004)}]{wheatland2004}
{Wheatland}, M.~S., 2004.
\newblock {A Bayesian Approach to Solar Flare Prediction}.
\newblock \emph{Astrophysical Journal}, \textbf{609}, 1134--1139.
\newblock 10.1086/421261.

\bibitem[{{Wiegelmann} and {Sakurai}(2012)}]{Wiegelmann:2012}
{Wiegelmann}, T., and T.~{Sakurai}, 2012.
\newblock {Solar Force-free Magnetic Fields}.
\newblock \emph{Living Reviews in Solar Physics}, \textbf{9}, 5.
\newblock 10.12942/lrsp-2012-5.

\bibitem[{Woodcock(1976)}]{Woodcock1976}
Woodcock, F., 1976.
\newblock The Evaluation of Yes/No Forecasts for Scientific and Administrative
  Purposes.
\newblock \emph{Monthly Weather Review}, \textbf{104}(10), 1209--1214.
\newblock \url{10.1175/1520-0493(1976)104<1209:TEOYFF>2.0.CO;2}.

\bibitem[{Xu et~al.(2019)Xu, Cisewski-Kehe, Green, and Nagai}]{Xu:2018xnz}
Xu, X., J.~Cisewski-Kehe, S.~B. Green, and D.~Nagai, 2019.
\newblock {Finding cosmic voids and filament loops using topological data
  analysis}.
\newblock \emph{Astron. Comput.}, \textbf{27}, 34--52.
\newblock 10.1016/j.ascom.2019.02.003.

\bibitem[{Yang et~al.(2013)Yang, Lin, Zhang, and Mao}]{yang2013}
Yang, X., G.~Lin, H.~Zhang, and X.~Mao, 2013.
\newblock Magnetic Nonpotentiality in Photospheric Active Regions as a
  predictor of solar flares.
\newblock \emph{Astrophysical Journal}, \textbf{774}(2), L27.
\newblock 10.1088/2041-8205/774/2/l27.

\bibitem[{Yu et~al.(2010)Yu, Huang, Wang, Cui, Hu, and Zhou}]{yu2010}
Yu, D., X.~Huang, H.~Wang, Y.~Cui, Q.~Hu, and R.~Zhou, 2010.
\newblock Short-term Solar Flare Level Prediction Using a Bayesian Network
  Approach.
\newblock \emph{Astrophysical Journal}, \textbf{710}(1), 869.
\newblock 10.1088/0004-637x/710/1/869.

\bibitem[{Yuan et~al.(2010)Yuan, Shih, Jing, and Wang}]{yuan2010}
Yuan, Y., F.~Shih, J.~Jing, and H.~Wang, 2010.
\newblock Solar flare forecasting using sunspot-groups classification and
  photospheric magnetic parameters.
\newblock \emph{Proceedings of the International Astronomical Union},
  \textbf{6}, 446 -- 450.
\newblock 10.1017/S1743921311015742.

\bibitem[{Zheng et~al.(2019)Zheng, Li, and Wang}]{Zheng2019}
Zheng, Y., X.~Li, and X.~Wang, 2019.
\newblock Solar Flare Prediction with the Hybrid Deep Convolutional Neural
  Network.
\newblock \emph{Astrophysical Journal}, \textbf{885}(1), 73.
\newblock 10.3847/1538-4357/ab46bd.

\bibitem[{Zomorodian(2012)}]{Zomorodian12}
Zomorodian, A., 2012.
\newblock Topological Data Analysis, vol.~70 of \emph{Advances in Applied and
  Computational Topology}.
\newblock American Mathematical Society, Providence.
\newblock 10.1090/psapm/070.

\end{thebibliography}
   

\Online

\begin{appendix} 
\section{Artificial Neural Network: Design and Implementation}
\label{sec:appendix_ann}

To evaluate the different feature sets, we designed a standard
feedforward neural network using \textsc{Keras} \citep{Chollet2015}
with six densely connected layers. The input layer size is variable
depending on the size of the feature set; the output layer contains
two neurons corresponding to the two classes---flaring and
non-flaring.  The four intermediate layers contain 12, 24, 16 and 8
neurons respectively, when counting from the direction of the input to
the output layer.  To prevent over-fitting, a Ridge Regression
regularization with a penalty factor of 0.01 is used at each layer
that limits the $L_{2}$ sum of all the weights.   
\revisiona{For the study reported here, we designed the neural network in a 
trial-and-error process that balanced complexity of representation
against training and testing time.  In future work, we play to
formalize this approach of fine-tuning the model hyperparameters
(learning rates, regularization parameters, loss weights, etc.) using
a separate validation and a testing set (as opposed to only using a
testing set).  The validation set can be used for determining the
optimal hyperparameter values, with the testing set used for a final
evaluation of the model.}
 
The layout of the network is shown in the figure below; the network parameters are 
summarized in Tbl.~\ref{tab:ann_parameters}.

\tikzset{%
	every neuron/.style={
		circle,
		draw,
		minimum size=1cm
	},
	neuron missing/.style={
		draw=none, 
		scale=4,
		text height=0.333cm,
		execute at begin node=\color{black}$\vdots$
	},
}

\vspace{10mm}
\begin{figure}[h]
\centering
\includegraphics[width=0.5\textwidth]{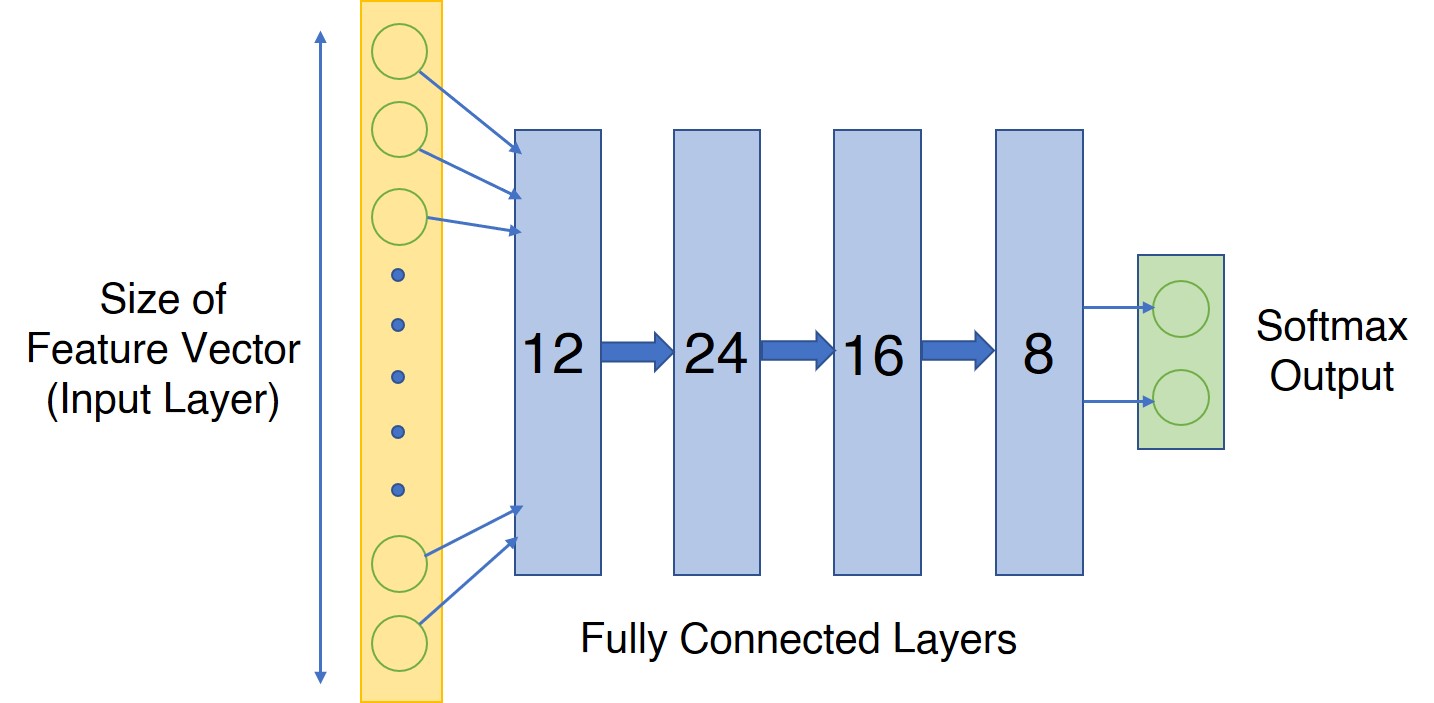}
\label{fig:ann}
\caption{Representation of the feed-forward Artificial Neural Network
  (ANN) model used in this work.  The number of neurons in the input
  layer is equal to the size of the feature set being used.  The
  output layer has two neurons---one each for the flaring and
  non-flaring classes.}

\end{figure}

We used the standard back-propagation algorithm built
into \textsc{Keras} to train the weights of the ANN, which are
initialized using the Glorot uniform initializer \citep{Glorot2010} at
the beginning of the training process. Since our problem is a two
category classification one, we chose a weighted binary cross-entropy
function to measure the training loss between the model output and the
true target that is propagated backwards to update the network
weights.  Over a training instance for $N$ data samples over $C$
classes (here $C=2$), this loss function is given by
\begin{equation}
L = \sum_{i=1}^{N} \sum_{c=1}^{C} w_{c}y^{'}_{ic}log p(y_{ic}),
\end{equation}
\noindent where $p(y_{ic})$ represents the target prediction
probability for class $c$ of the $i^{th}$ data point, whose real
target value is $y^{'}_{ic}$.  The loss function class weights $w_{c}$
in this formula determine the penalty of incorrectly predicting the
probability for the associated classes $c$.  As demonstrated in
Section~\ref{sec:data}, our dataset is highly unbalanced, with the
non-flaring class over-represented by two orders of magnitude.  By
choosing $w_{no-flare}:w_{flare} = 1:100$, we over-penalize the
minority class (flaring magnetograms) to offset the effect of its
size.  This is one of the many ways to mitigate imbalance in training
sets, and has been used in the flare-prediction
literature \citep{Bobra:2015fn, Nishizuka:2018}.  Finally, we use the
Adagrad optimizer \citep{Duchi2011} to perform the weight update during 
the backpropagation phase. An extension of the Stochastic Gradient Descent algorithm, the
Adagrad optimizer adapts the learning rate to the individual parameters, so
that more sensitive parameters which significantly affect the output 
are assigned a lower learning rate, whereas less sensitive parameters
are updated with a greater learning rate. (Our initial choice was the 
Adam optimizer \citep{Kingma2014}, but we observed jitters in the prediction
scores after convergence when using it.)  
We used the default parameters for the \textsc{PyTorch} implementation with the 
learning rate multiplier set to $5 \times 10^{-4}$. 

\begin{table}
	\centering
	\begin{tabular}{c c}
		\hline
		\textbf{Model Parameters} & \textbf{Settings} \\
		\hline
		Total hidden layers & 4 \\
		Hidden layer activation function & Rectified Linear Unit \\
		Output layer activation function & Softmax \\
		Weight regularization & L2 with a penalty of $10^{-2}$ \\
		Training batch size & \revisiona{64} \\
		Loss function & Weighted Binary Cross-Entropy (weights: $w_{no-flare} = 1$, \revisiona{$w_{flare} = 100$})\\
		Optimization Method & \revisiona{Adagrad} (Parameters: Learning rate: \revisiona{$5 \times 10^{-4}$})\\
		\revisiona{Number of Epochs} & \revisiona{50} \\
		\hline
	\end{tabular}
	\caption{Parameter settings for the Artificial Neural Network (ANN) model implementation.}
	\label{tab:ann_parameters}
\end{table}

\revisiona{All  experiments were run on a NVIDIA TITAN V graphics processing
unit (NVIDIA Driver Version 440.31, CUDA Version 10.2).  Across all
feature sets, the training time per epoch was approximately 22s.  For
a total of 50 epochs, the total training time then amounted to
approximately 19 minutes. The average validation time across all the
feature sets was approximately 13s.}

\section{\revisiona{Alternative Evaluation Metrics}}
\label{sec:appendix_results_all_metrics}

\revisiona{Here, we report the performance
of the various feature sets using some other standard metrics.
Comparing the forecast against the actual event, we populate the
contingency table values: true positives (TP), false positives (FP),
true negatives (TN) and false negatives (FN). Using these four
classes, we compute the five additional metrics for each experiment:
Accuracy, Precision, Recall, the F-1 Score (F1) and the Heidke Skill
Score (HSS):}
\begin{equation}
Accuracy = \frac{TP + TN}{TP + TN + FP + FN}
\end{equation}
\begin{equation}
Precision = \frac{TP}{TP + FP}
\end{equation}
\begin{equation}
Recall = \frac{TP}{TP + FN}
\end{equation}
\begin{equation}
F1 = \frac{2 \times Precision \times Recall}{Precision + Recall}
\end{equation}
\begin{equation}
HSS = \frac{2(TP \times TN - FP \times FN)}{(TP + FN)(FN + TN) + (TP + FP)(FP + TN)}
\end{equation}

\revisiona{For a detailed explanation of these metrics,  
refer to \cite{Bobra:2015fn} (note that the HSS$_{2}$ variant of the
Heidke Skill Score from \cite{Bobra:2015fn} is used here).  The mean
metric scores and the mean score improvements along with the standard
deviations over the baseline the geometry and topology experiments are
summarized in Tbl.~\ref{tab:geometry_results_all_metrics}
and Tbl.~\ref{tab:topology_results_all_metrics} respectively. The trends
for these metrics match the trends in the TSS scores described in
Section~\ref{sec:results}.}

\begin{table}[H]
	\centering
	\begin{tabular}{c c c c c c}
		\hline
		Feature Set & Accuracy & Precision & Recall & F-1 Score & HSS \\
		\hline
		\hline
		SHARP &	0.89 $\pm$ 0.01 & 0.09 $\pm$ 0.01 & 0.79 $\pm$ 0.06 & 0.15 $\pm$ 0.01 & 0.13 $\pm$ 0.01\\ 
		Geometry & 0.88 $\pm$ 0.01	& 0.08 $\pm$ 0.01 & 0.79 $\pm$ 0.06 & 0.14 $\pm$ 0.01 & 0.12 $\pm$ 0.01 \\
		SHARP + Geometry & 0.90  $\pm$ 0.01 & 0.09 $\pm$ 0.01 & 0.82 $\pm$ 0.06 & 0.17 $\pm$ 0.01 & 0.15 $\pm$ 0.01 \\
		\hline
		Geometry Improvement & -0.01 $\pm$ 0.00 & -0.01 $\pm$ 0.00 & 0.00 $\pm$ 0.03 & -0.01 $\pm$ 0.01 & -0.02 $\pm$ 0.01 \\
		SHARP + Geometry Improvement & 0.01 $\pm$ 0.00 & 0.01 $\pm$ 0.00 & 0.03 $\pm$ 0.02 & 0.01 $\pm$ 0.00 & 0.01 $\pm$ 0.00 \\
		\hline
	\end{tabular}
	\caption{\revisiona{Performance evaluations of the various
		feature sets for the geometry experiments using
		various metrics. The top three rows report the mean
		metric score (with the standard deviation) across ten
		different training/testing sets, while the bottom two
		rows report the mean improvement with the standard
		deviation of the metric scores with respect to the
		SHARP feature set.}}
	\label{tab:geometry_results_all_metrics}
\end{table}

\begin{table}[H]
	\centering
	\begin{tabular}{c c c c c c}
	\hline
	Feature Set & Accuracy & Precision & Recall & F-1 Score & HSS \\
	\hline
	\hline
	SHARP &	0.89 $\pm$ 0.01 & 0.09 $\pm$ 0.01 & 0.79 $\pm$ 0.06 & 0.15 $\pm$ 0.01 & 0.13 $\pm$ 0.01\\ 
	Topology & 0.90 $\pm$ 0.01 & 0.10 $\pm$ 0.01 & 0.81 $\pm$ 0.06 & 0.18 $\pm$ 0.02 & 0.16 $\pm$ 0.02 \\
	SHARP + Topology & 0.90 $\pm$ 0.01 & 0.09 $\pm$ 0.01 & 0.83 $\pm$ 0.05 & 0.17 $\pm$ 0.02 & 0.15 $\pm$ 0.02 \\
	\hline
	Topology Improvement & 0.01 $\pm$ 0.01 & 0.01 $\pm$ 0.01 & 0.02 $\pm$ 0.05 & 0.02 $\pm$ 0.01 & 0.02 $\pm$ 0.01 \\
	SHARP + Topology Improvement & 0.01 $\pm$ 0.01 & 0.01 $\pm$ 0.00 & 0.04 $\pm$ 0.04 & 0.01 $\pm$ 0.01 & 0.02 $\pm$ 0.01 \\
	\hline
	\end{tabular}
	\caption{\revisiona{Performance evaluations of the various
		feature sets for the topology experiments using
		various metrics. The top three rows report the mean
		metric score (with the standard deviation) across ten
		different training/testing sets, while the bottom two
		rows report the mean improvement with the standard
		deviation of the metric scores with respect to the
		SHARP feature set.}}
	\label{tab:topology_results_all_metrics}
\end{table}

\section{Feature Ranking}
\revisiona{Evaluating the model performance scores using different
	feature sets is one way of determining how effective the engineered
	features are in providing an accurate model prediction.  An
	alternative evaluation strategy is to compute the Fisher score
	\citep{Gu2012} individually for each feature, as done in
	\cite{Bobra:2015fn}.  A method of univariate feature ranking, the
	Fisher score (or the F-score) determines the ability of the feature
	to separate the distributions of the classes in the dataset. For a
	dataset with two labels, the F-score of a feature $i$, as defined in
	\cite{Bobra:2015fn}, is given by}

\begin{equation}
F(i) = \frac{(\bar{x}_{i}^{+} - \bar{x}_{i})^{2} + (\bar{x}_{i}^{-} - \bar{x}_{i})^{2}}{\frac{1}{n^{+} - 1}\sum_{k=1}^{n^{+}}(x_{k,i}^{+} - \bar{x}_{i})^{2} + \frac{1}{n^{-} - 1}\sum_{k=1}^{n^{-}}(x_{k,i}^{-} - \bar{x}_{i})^{2}},
\end{equation}
\revisiona{where $\bar{x}_{i}$, $\bar{x}_{i}^{+}$ and
	$\bar{x}_{i}^{-}$ represent the average feature value over samples
	of the full dataset: $n^{+}$ samples belonging to the positive class
	and $n^{-}$ samples belonging to the negative class respectively.
	The numerator in the above equation represents the inter-class
	distance or separability, while the denominator computes the
	intra-class variance.  Thus, a feature with a smaller spread within
	each of the two classes and a higher separation between the two
	class means would generate a higher F-score.}

\begin{figure}[h]
	
	\centering \subfloat[Geometry vs. SHARPs]{
		\includegraphics[width=0.5\textwidth]{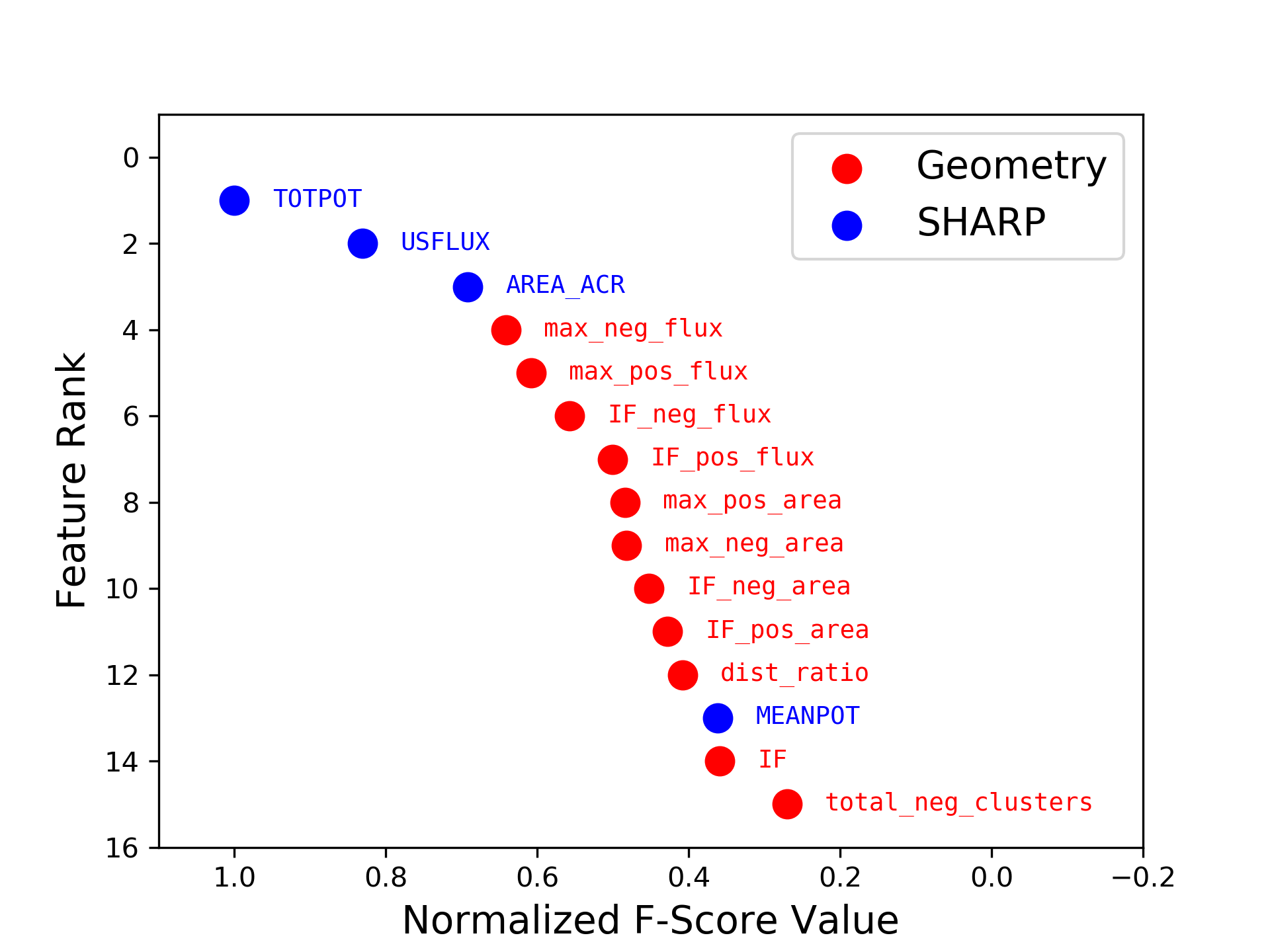}
	} \subfloat[Topology vs. SHARPs]{
		\includegraphics[width=0.5\textwidth]{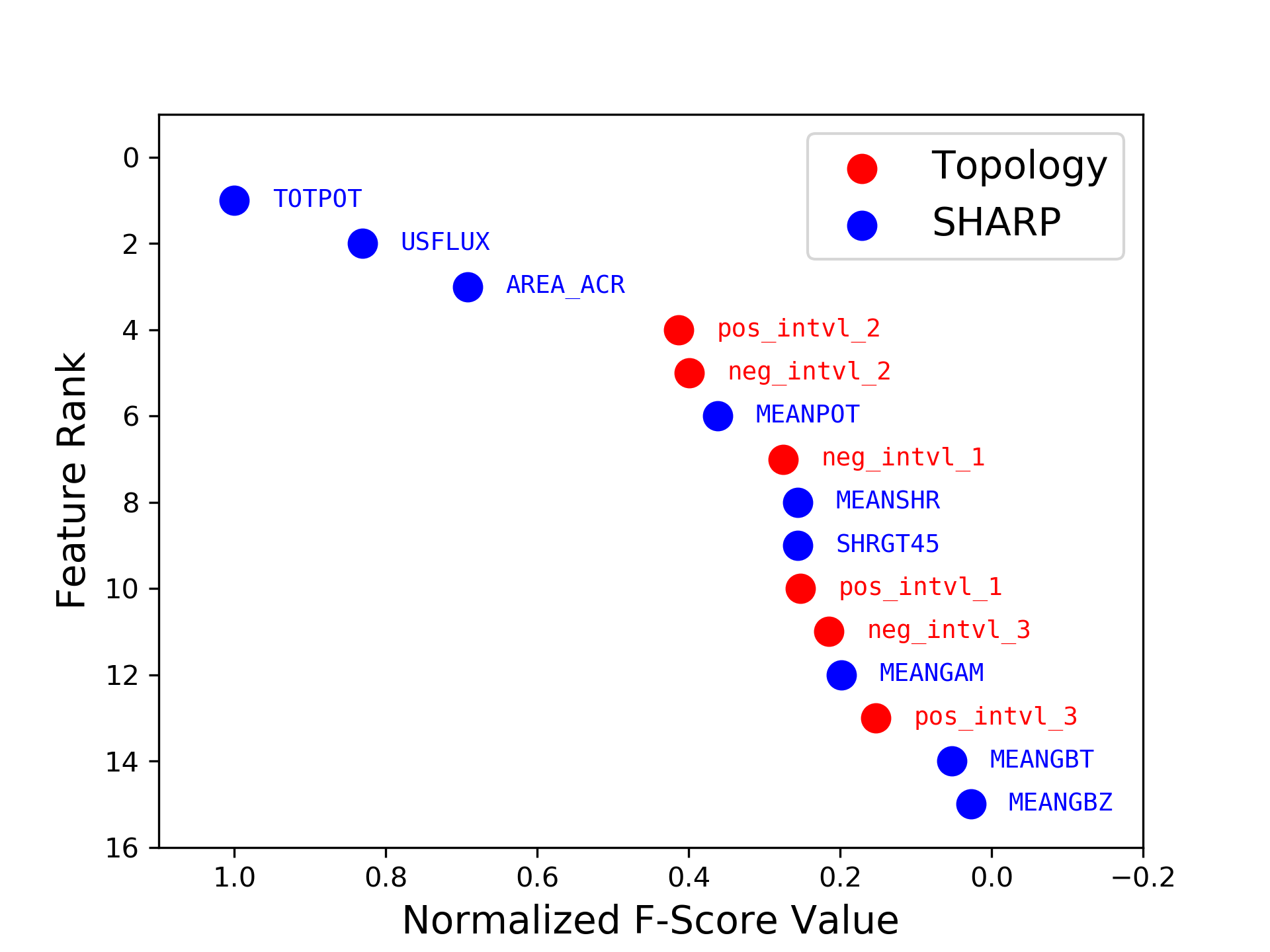}
	}
	\caption{\revisiona{Feature ranking, as determined via the
			Fisher score method, for the experiments reported in this
			paper. The x-axis is the normalized Fisher score and the 
			y-axis is the feature rank. Red identifies features based on 
			topology and geometry; blue indicates SHARPs features. 
			The features
			are plotted in descending order of the F-Score value on
			the x-axis and ascending order of the rank on the y-axis
			(both axes are inverted).  The topological features
			described by the formats {\tt pos\_intvl\_n} and
			{\tt neg\_intvl\_n} represent the number of holes for the
			$n^{th}$ magnetic flux threshold for the positive and
			negative polarities respectively.}}
	\label{fig:feature_ranking}
\end{figure}

\revisiona{For all the features in the SHARPs-plus-geometry and
	SHARPs-plus-topology combination feature sets from the
	geometry and the topology experiments, we compute the normalized
	F-score scaled with respect to the highest scoring feature. We use
	the {\tt f\_classif} functionality in the {\tt Python} {\tt
		Scikit-Learn} package \citep{Pedregosa2011} to compute the
	F-score, and report the top 15 ranking features. The results are
	shown in Figure~\ref{fig:feature_ranking}.}
\revisiona{In all experiments, the top three ranking features---total
	photospheric energy density ({\tt TOTPOT}), total unsigned flux ({\tt USFLUX})
	and the total area of the active pixels ({\tt AREA\_ACR})---belong to the
	SHARPs feature set.  In the geometry-based experiments (Figure
	\ref{fig:feature_ranking}(a)), all but one of the top 15 properties belong to
	the geometry feature set.  The higher-ranking of these pertain to
	the magnetic flux: the largest area ({\tt max\_pos\_flux, max\_neg\_flux})
	and the polarities belonging to the MIP ({\tt IF\_pos\_flux,
		IF\_neg\_flux}), followed by the areas of the respective polarities
	(geometry features with the ``{\tt \_area}" suffix).  In the topology-based
	experiments (Figure \ref{fig:feature_ranking}(b)), the feature ranking is more mixed.
	Six of the top 15 features belong to the topology feature set: the
	number of live holes at the magnetic flux values $\pm 263.16$ $G$
	({\tt pos\_intvl\_1, neg\_intvl\_1}), $\pm 789.47$ $G$ ({\tt pos\_intvl\_2,
		neg\_intvl\_2}) and $\pm 1315.79$ $G$ ({\tt pos\_intvl\_3, neg\_intvl\_3}).  In
	terms of univariate feature ranking, the geometry (nine of the top
	15 features) and topology features (six of the top 15) perform well
	individually when compared with SHARPs features.}

\revisiona{These scores demonstrate the predictive power of {\sl
		individual} features in terms of their ability to discriminate the
	flaring and non-flaring datasets.  However, this cannot be directly
	correlated with the TSS scores of the different feature sets
	described above. The Fisher score is a univariate feature ranking
	method and does not take into account the correlation between the
	different features.  We leave this investigation of the correlation
	within and between the various feature sets as future work.}

\label{sec:appendix_feature_ranking}

\end{appendix}

\end{document}